\UseRawInputEncoding

\documentclass[aps,prx,twocolumn,superscriptaddress,noeprint,longbibliography]{revtex4-2}

\usepackage{amsmath,amsthm,amssymb,amsfonts}
\usepackage{float}
\usepackage{physics}

\usepackage{graphics}
\usepackage[utf8]{inputenc}
\usepackage{tikz}
\usetikzlibrary{positioning}
\usepackage{graphicx, subfigure}
\usepackage{float}
\usepackage{comment}
\usepackage{hyperref}
\usepackage{tcolorbox}
\usepackage[inline]{enumitem}
\usepackage{amsmath,amsthm,amssymb,amsfonts}
\usepackage{physics}
\usepackage{stackengine,graphicx}
\usepackage{graphicx}
\usepackage{dcolumn}
\usepackage{bm}
\usepackage{hyperref}
\usepackage{array}
\usepackage{multirow}
\newcolumntype{P}[1]{>{\centering\arraybackslash}p{#1}}

\hypersetup{
	colorlinks = true,
	citecolor = {blue}
	}
\usepackage[mathlines]{lineno}
\UseRawInputEncoding
\graphicspath{ {./images/} }

\newcommand{\lht}[1]{\textcolor{orange}{LHT: \texttt{#1}}}

\begin{document}
 
\preprint{APS/123-QED}

\title{Multi-magnon quantum many-body scars from tensor operators}
\author{Long-Hin Tang}
\email{lhtang@bu.edu}
\affiliation{Department of Physics, Boston University, MA 02215, USA}
\author{Nicholas O'Dea}
\affiliation{Department of Physics, Stanford University, Stanford, CA 94305, USA}
\author{Anushya Chandran}
\affiliation{Department of Physics, Boston University, MA 02215, USA}

\date{\today}

\begin{abstract}
We construct a family of three-body spin-1/2 Hamiltonians with a super-extensive set of infinitely long-lived multi-magnon states. 
A magnon in each such state carries either quasi-momentum zero or fixed $p_0\neq 0$, and energy $\Omega$. These multi-magnon states provide an archetypal example of quantum many-body scars: they are eigenstates at finite energy density that violate the eigenstate thermalization hypothesis, and lead to persistent oscillations in local observables in certain quench experiments.
On the technical side, we demonstrate the systematic derivation of scarred Hamiltonians that satisfy a restricted spectrum generating algebra using an operator basis built out of irreducible tensor operators. This operator basis can be constructed for any spin, spatial dimension or continuous non-Abelian symmetry that generates the scarred subspace.
\end{abstract}
\maketitle

\section{Introduction}\label{sec:introduction}
Dynamics based on elementary local moves need not be ergodic. Examples abound in frustrated magnets~\cite{dolinsek_broken_2008,lee_frustration-induced_2021}, kinetically constrained systems~\cite{lan_quantum_2018}, spin and molecular glasses~\cite{baldwin_clustering_2017,garrahan_aspects_2018,rademaker_slow_2020}, and systems with multipole conservation laws~\cite{sala_ergodicity_2020,khemani_localization_2020}. In isolated quantum systems, this breakdown of ergodicity can be detected through athermal eigenstates that violate the Eigenstate Thermalization Hypothesis (ETH)~\cite{jensen_statistical_1985, deutsch_quantum_1991, srednicki_chaos_1994, rigol_thermalization_2008, dalessio_quantum_2016}.

Recent experiments in Rydberg simulators~\cite{bernien_probing_2017,celi_emerging_2020,bluvstein_controlling_2021} have brought a particular class of such Hamiltonians into prominence, namely those with \emph{quantum many-body scars}~\cite{heller_bound-state_1984, shiraishi_systematic_2017,turner_weak_2018,moudgalya_exact_2018,schecter_weak_2019,choi_emergent_2019, khemani_signatures_2019,iadecola_quantum_2019,lin_exact_2019,ok_topological_2019,hudomal_quantum_2020,shibata_onsagers_2020,srivatsa_quantum_2020,mcclarty_disorder-free_2020,hart_compact_2020,moudgalya_rsga_2020, mark_unified_2020,mark__2020,alhambra_revivals_2020,lee_exact_2020,iadecola_quantum_2020,zhao_quantum_2020, odea_tunnels_2020,pakrouski_many-body_2020,ren_quasisymmetry_2021,chertkov_motif_2021,martin_scar_2021,langlett_rainbow_2021,van_voorden_disorder_2021,yao_quantum_2021,banerjee_quantum_2021,jeyaretnam_quantum_2021,zhao_orthogonal_2021,desaules_hypergrid_2021,sharma_multimagnon_2022,jepsen_catching_2021,you_quantum_2022,schindler_exact_2022}. Generic initial states thermalize under time evolution by scarred Hamiltonians because almost all eigenstates satisfy the ETH. Quantum many-body scars are the special atypical, highly excited eigenstates that do not satisfy the ETH. The number of scarred eigenstates usually scales as a polynomial of the system volume. If one prepares a state restricted to the scarred manifold, then it exhibits persistent coherent oscillations without thermalization.\par 

A unifying feature of the Hamiltonian relevant to the Rydberg experiment~\cite{bernien_probing_2017}  and a subclass~\footnote{Scars can come in isolation without any clear structure or fixed energy gaps. Alternatively, scars can be generated by multiple independent ladder operators, each adding different energies, so that the scars do not form a simple tower.} of other scarred Hamiltonians in the literature~\cite{moudgalya_exact_2018,schecter_weak_2019, moudgalya_rsga_2020,mark_unified_2020,odea_tunnels_2020,pakrouski_many-body_2020,ren_quasisymmetry_2021} is that the scarred states (approximately) (i) form a harmonic ladder (or tower) with frequency ${\Omega}$, and (ii) are generated by repeated application of a quasiparticle annihilation operator (or ladder operator) $Q^-$ on a simple low-entanglement base eigenstate $\ket{\psi_0}$: 
\begin{equation}\label{eq:tower}
    \ket{\psi_n}=\big(Q^-\big)^{n}\ket{\psi_0},\quad n\geq0\quad.
\end{equation}
That is, the Hamiltonian ${H}$ satisfies a spectrum generating algebra (SGA)~\cite{mark_unified_2020,moudgalya_rsga_2020} in the scarred manifold $W$ spanned by the states $\ket{\psi_n}$,
\begin{equation}\label{eq:SGA0}
    \big[H,Q^-\big]W=-\Omega Q^-W.
\end{equation}
Note that the majority of the literature uses a creation operator $Q^+$ to generate the scar tower.\par

\begin{figure}[t]
\includegraphics[width=\linewidth]{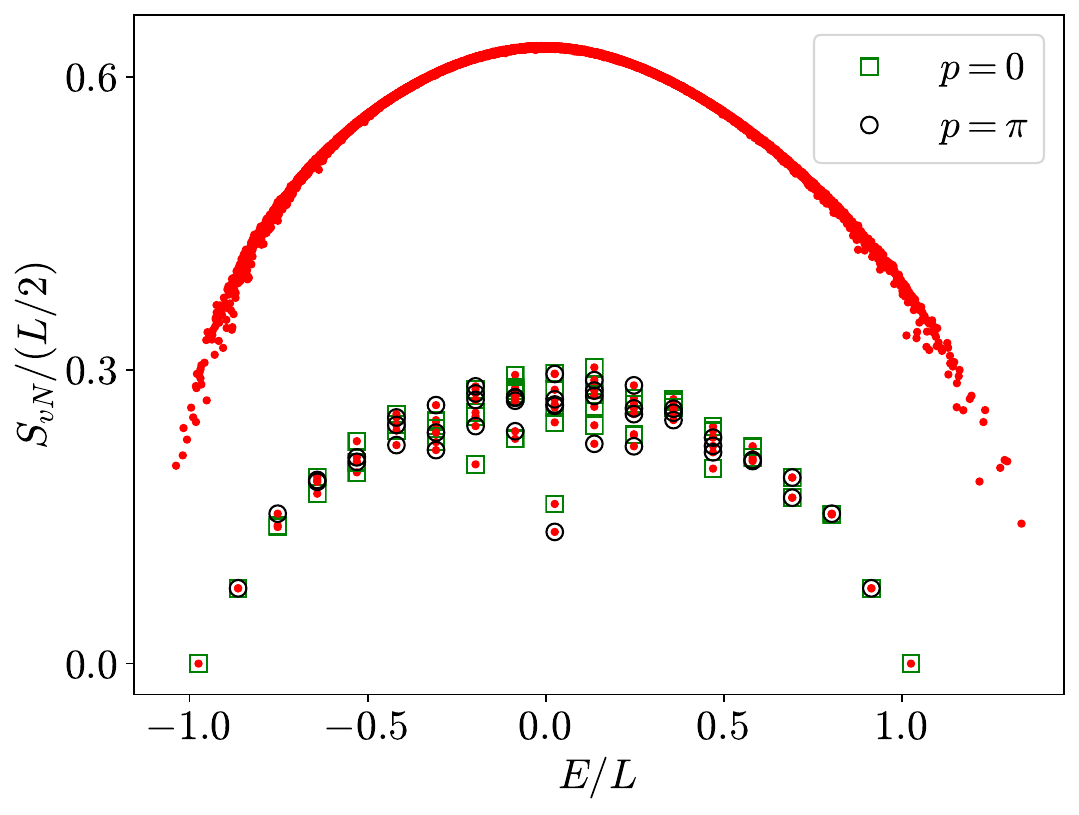}
		\caption{The half-chain entanglement entropy density plotted against the energy density of each eigenstate of ${H_{\pi}}$ (given by Eq.~\eqref{eq:multipimagnon}) in the ${p=0}$ (with 55 scarred states marked by squares) and ${p=\pi}$ (with 45 scarred states marked by circles) momentum sectors at ${L=18}$. Other than these outlying states, the entropy density is a smooth function of the energy density, indicating that the model is otherwise thermalizing. }
\label{fig:piSE}
	
\end{figure}

Coherent states in $W$ oscillate in time and exhibit perfect periodic revivals. If the operator ${Q^-}$ is a \emph{sum} of single-site operators and the state ${\ket{\psi_0}}$ is a product state, then the coherent state is also a product state which can be easily prepared in quench experiments. The bipartite entanglement entropy of the scarred eigenstates also scales at most logarithmically with the subsystem size~\cite{moudgalya_entanglement_2018,odea_tunnels_2020}, so that the scars can be numerically identified as the low-entanglement outliers, as in Fig.~\ref{fig:piSE}.\par 

In this manuscript, we extend the theory of exact quantum many-body scars in two ways. Consider quasiparticle creation and annihilation operators that satisfy the SU(2) algebra
\begin{equation}\label{eq:su2op}
    Q^+=\big(Q^-\big)^{\dagger},\quad Q^z=\big[Q^+,Q^-\big]/2.
\end{equation}
Our first result (Sec.~\ref{sec:tensorrsga}) is that a Hamiltonian ${H}$ that satisfies Eq.~\eqref{eq:SGA0} has a natural representation in the ${Q}$-SU(2) spherical tensor basis
\begin{equation}\label{eq:introtensorexpand}
    H=\sum_{k=0}\sum_{\phi_k}\sum_{q=-k}^{k}c_q^{\phi_k}T_{(k),\phi_k}^{q}.
\end{equation}
Above, ${q}$ and ${k}$ are the component and rank of a spherical tensor ${T^{q}_{(k),\phi_k}}$, ${\phi_k}$ labels the representation, and ${c^{\phi_k}_q}$ are complex coefficients. The spherical tensor basis is advantageous in two respects. First, the constraint equations for these coefficients are simplified by the property that ${\big[Q^{\pm},T^{q}_{(k),\phi_k}\big]\propto T^{q\pm1}_{(k),\phi_k}}$, so that commutation with ${Q^-}$ is a permutation in the spherical tensor basis. Further, if ${\ket{\psi_0}}$ is a highest-weight state with fixed ${z}$-magnetization, the equations for different ${k}$ and ${q}$ components separate. Each of these equations involve smaller numbers of coefficients (Table~\ref{fig:dimension}), making analytical solutions feasible. Second, the number of nontrivial equations is set by the maximum ${q}$ and the corresponding ${k}$ in Eq.~\eqref{eq:introtensorexpand} (see Eq.~\eqref{eq:nestedcom}). By restricting these values in the expansion, one may explore degrees of freedom in operator space more efficiently as the range of the operators increases.


The Hamiltonians ${H}$ constructed using the spherical tensor formalism need not annihilate the states in the scarred manifold term-by-term. They may therefore lie beyond the projector-embedding formalism introduced by Shiraishi and Mori~\cite{shiraishi_systematic_2017} (SM) and used in Refs.~\cite{shiraishi_connection_2019, srivatsa_quantum_2020, ok_topological_2019} to construct scarred models. In addition, ${H}$ can be naturally classified by a \emph{restricted spectrum generating algebra} (RSGA), as formulated in Ref.~\cite{moudgalya_rsga_2020}.\par 

Our second result (Sec.~\ref{sec:magnonscar}) is that Hamiltonians of the form in Eq.~\eqref{eq:introtensorexpand} can have multi-magnon states as scarred states. Specifically, using up to three-body terms, we design periodic 1d spin-${s}$ Hamiltonians with any number of ${p=0}$ magnons and one or more fixed ${p_0\neq0}$ magnons atop the fully polarized state, where ${p}$ is the (quasi)momentum of a magnon. We focus on the ${s=1/2}$ case in the main text and discuss $s>1/2$ in Appendix~\ref{sec:spinsgen}. 

The multi-magnon states on $L$-site chains are of the form 
\begin{equation}\label{eq:multimagnonstate}
\begin{aligned}
    \ket{\psi_{n,N}(p_0)}&=\big(S^-\big)^n\big(Q^-(p_0\neq0)\big)^N\bigotimes_{i=0}^{L-1}\ket{\uparrow}_{i},\\
    Q^-(p)&=\sum_{i=0}^{L-1} e^{-ipr_i}S^-_{i},\quad Q^-(p=0)\equiv S^-.
\end{aligned}
\end{equation}
A straightforward computation (Sec.~\ref{sec:SManni}) identifies two projectors that locally annihilate the multi-magnon states with any number  of $p=0$ magnons and up to one magnon with a given momentum $p_0$~\footnote{As the SM Hamiltonians conserve total z-magnetization, all single magnon states are also eigenstates.}. One of the two projectors furthermore annihilates states with multiple $p_0$-magnons. The corresponding SM Hamiltonian thus has a \emph{scar pyramid} with (degenerate) scarred manifolds that are equally spaced in energy in equally spaced quasi-momentum sectors. 

Sec.~\ref{sec:generalHam} derives scarred Hamiltonians that lie beyond the SM formalism. In particular, the family of Hamiltonians involving next-nearest neighbor Dzyaloshinskii-Moriya (DM) interaction terms and rank-2 tensors
\begin{equation}\label{eq:tensorresult}
    H_{\text{A}}=\sum_{\mu=x,y,z}J_{\mu}^{\text{DM2}}H^{\text{DM}}_2(\mu)+\sum_{\lambda=1}^{5}J_{\lambda}^{\text{RT}}H^{\text{RT}}(\lambda),
\end{equation}
annihilate multi-magnon states with any number of ${p=0}$ and ${p_0=\pi}$ magnons. These scarred states are identified as low-entanglement outliers in Fig.~\ref{fig:piSE}. In quench experiments, these scarred states lead to perfect and persistent revivals of fidelity for a family of initial period-2 product states. In the presence of nearest-neighbor Heisenberg or Dzyaloshinskii-Moriya interaction terms, the multi $p_0=\pi$ magnon states are no longer perfect scars. We study the concomitant decay of the fidelity oscillations in Secs.~\ref{sec:pipi} and \ref{sec:deformpipi}.\par

There are a few examples of scar pyramids generated by multiple ladder operators in the literature. These differ from our construction in detail: the ladder operators are either multi-site or non-local~\cite{iadecola_quantum_2020,mark_unified_2020,mark__2020}, or are derived from the multiple ladder operators of non-Abelian symmetries like SU(3)~\cite{ren_quasisymmetry_2021,odea_tunnels_2020, ren_deformed_2022}. The base states in these and other~\cite{schecter_weak_2019, mark__2020, moudgalya_rsga_2020} constructions are furthermore the highest/lowest weight states under the Casimir operator ${S^2}$, unlike the scarred states with ${N=1}$. More strikingly, the scarred states for ${1<N,n<L-1}$ are neither eigenstates of the Casimir operators ${S^2}$ nor ${Q^2(p)}$; the analogous situation for a single scarred tower is only realized in the Affleck-Kennedy-Lieb-Tasaki (AKLT) model ~\cite{mark_unified_2020,moudgalya_exact_2018,moudgalya_entanglement_2018} and its q-deformed and higher-rank symmetry generalizations~\cite{odea_tunnels_2020}.\par

Our construction applies mutatis mutandis to cases where the operator ${Q^-}$ is not the lowering operator of the bare spin-SU(2) algebra. In Refs.~\cite{mark__2020,schecter_weak_2019}, scarred towers were generated by creating ${p_0=\pi}$ (bi)magnon excitations on top of a fully polarized state ${\ket{\psi_0}}$, i.e. ${\ket{\psi_n}=\big(Q^-\big)^n\ket{\psi_0}}$ with either ${Q^-=\sum_{i}(-1)^iS^-_{i}}$ or ${Q^-=\sum_{i}(-1)^i(S^-_{i})^2/2}$. In Appendix~\ref{sec:multipole}, we show that such scar towers are embedded by a linear combination of the corresponding ${Q}$-SU(2) tensor operators. We also present a method for generating complete bases of tensor operators associated with a chosen continuous non-Abelian symmetry, and apply it to SU(3). \par  

In what follows, we briefly review (Sec.~\ref{sec:review}) the RSGA conditions and a symmetry-based framework \cite{odea_tunnels_2020} for obtaining scarred Hamiltonians. Then, after presenting the tensor operator formalism (Sec.~\ref{sec:tensorrsga}) and a family of models with magnon scarred states (Sec.~\ref{sec:magnonscar}), we numerically simulate two particular models from the family (Sec.~\ref{sec:simulation}). Generalizations and technical details are relegated to the appendices.

\section{The Symmetry-based framework and RSGA}\label{sec:review}

A ladder operator ${Q^-}$ and a Hamiltonian ${H}$ which satisfies the SGA in Eq.~\eqref{eq:SGA0} can be furnished from generators of Lie groups \cite{pakrouski_many-body_2020, pakrouski_group_2021, odea_tunnels_2020, ren_quasisymmetry_2021, ren_deformed_2022}. For ${H}$ to be a scarred, it must break the symmetry associated with the group. We make extensive use of one such framework \cite{odea_tunnels_2020} which employs this principle.\par 

A key element of this framework is a set of spectrum generating operators (SGOs) furnished by the generators of the Lie group ${G}$. To be specific, in the case that ${G=}$SU(2), the associated set of SGOs ${\{Q^+,Q^-,Q^z=\big[Q^+,Q^-\big]/2\}}$ satisfy the usual commutation relations
\begin{equation}
\big[Q^z,Q^\pm\big]=\pm Q^\pm.
\end{equation}
Then, a tower of target states ${\ket{\psi_n}}$ in the form of Eq.~\eqref{eq:tower} can be embedded as scars in a model with the following three components:
\begin{equation}
    H=H_{\text{sym}}+H_{\text{SG}}+H_{\text{A}}.
\end{equation}
\begin{enumerate}
    \item ${H_{\text{sym}}}$ is a SU(2)-symmetric term that contains multiplets of degenerate eigenstates which transform as irreducible representations (irreps) of the symmetry group SU(2). The multiplets are labelled by their eigenvalues under ${Q^z}$ and the Casimir operator ${Q^2=\{Q^+,Q^-\}/2+(Q^z)^2}$.

    \item  The degeneracy of these multiplets is lifted to equally spaced towers by the spectrum generating term ${H_{\text{SG}}=\Omega Q^{z}}$.

    \item Finally, the symmetry-breaking term ${H_{\text{A}}}$ preserves the target states as scars by annihilating them
    \begin{equation}\label{eq:annhilator}
        H_{\text{A}}\ket{\psi_n}=0\quad\forall n\geq0
    \end{equation}
    and connecting eigenstates across different symmetry sectors of ${H_{\text{sym}}}$ so that the rest of the energy spectrum is generic and thermalizing.
\end{enumerate}
A base state ${\ket{\psi_0}}$ need not be chosen to be an eigenstate of ${Q^2}$. For instance, in Eq.~\eqref{eq:multimagnonstate}, ${\{\ket{\psi_{0,N}(p_0)}\}}$ may be regarded as a set of base states, and those with ${1<N<L-1}$ are not eigenstates of ${S^2}$, which restricts the form of ${H_{\text{sym}}}$.
\par  

Scarred Hamiltonians satisfying a SGA can be further classified under the RSGA framework \cite{moudgalya_rsga_2020}. Suppose
\begin{equation}\label{eq:comh}
\begin{aligned}
    H_0&\equiv H,\quad H_{n+1}\equiv\big[H_{n},Q^-\big]\quad \forall n\geq0.
\end{aligned}
\end{equation}
Given ${\ket{\psi_0}}$ and ${Q^-}$, a Hamiltonian ${H}$ is said to exhibit a restricted spectrum generating algebra of order ${M}$ (RSGA-${M}$) if it satisfies 
\begin{equation}\label{eq:rsgacon}
        \begin{aligned}
    &\text{(i) }H\ket{\psi_0}=E_0\ket{\psi_0}\\
    &\text{(ii) }H_1\ket{\psi_0}=\big[H,Q^-\big]\ket{\psi_0}=-\Omega Q^-\ket{\psi_0}\\
    &\text{(iii) }H_n\ket{\psi_0}=0\quad\forall n\in\{2,3,\cdots M\}\\
    &\text{(iv) }H_n\begin{cases}
    \neq 0 \text{ if }n\leq M\\
    =0 \text{ if }n\geq M+1
    \end{cases}.
    \end{aligned}
\end{equation}

\section{Tensor operators as an operator basis}\label{sec:tensorrsga}
In this section, we introduce the irreducible tensor operator formalism for constructing scarred Hamiltonians. 

To be specific, consider a spin-s periodic chain with ${L}$ sites. We take ${Q^-=S^-=\sum_{i=0}^{L-1}S^-_{i}}$ as the total lowering operator, and ${\ket{\psi_0}=\ket{\psi_{0,N}(p_0)}}$  (defined in Eq.~\eqref{eq:multimagnonstate}) as base states. In Appendix~\ref{sec:multipole}, we present generalizations to other ${Q^-}$.\par

\subsection{Definitions and basic properties}\label{sec:defprop}
The essential algebraic properties of tensor operators~\cite{nielsen_spherical_2006,man_cartesian_2013} follow from their transformations under spatial rotations. Under infinitesimal rotations, a vector operator ${T^{\mu}}$ with component ${\mu=x,y,}$ or ${z}$ transforms as 
\begin{equation}
    \big[T^{\mu},S^{\nu}\big]=i\sum_{\lambda=x,y,z}\epsilon^{\mu\nu\lambda}T^{\lambda}
\end{equation}
where ${\epsilon^{\mu\nu\lambda}}$ is the Levi-Civita symbol.
%
Cartesian tensors are tensor products of vector operators
\begin{equation}\label{eq:reducible}
    T^{\mu_1\cdots\mu_n}=T_1^{\mu_1}\cdots T_{n}^{\mu_n}.
\end{equation}
They can be decomposed into spherical tensors that transform irreducibly under infinitesimal rotations. This leads to the defining commutation relations
\begin{equation}\label{eq:STalgebra}
\begin{aligned}
    \big[S^z,T_{(k)}^{q}\big] &=qT_{(k)}^{q}\\
    [S^{\pm},T_{(k)}^{q}]&=\sqrt{(k\mp q)(k\pm q+1)}T_{(k)}^{q\pm1}\\
\end{aligned}
\end{equation}
\begin{equation}\label{eq:rank}
\begin{aligned}
    \sum_{\mu=x,y,z}\big[S^{\mu},[S^{\mu},T_{(k)}^{q}]\big]=k(k+1)T_{(k)}^{q}
\end{aligned}
\end{equation}
where ${k}$ is the rank of the tensor and ${q=-k,-k+1,\cdots k-1, k}$ denotes the ${2k+1}$ components. Operators of different ${k}$ and ${q}$ are trace-orthogonal.\par 

Notice that the transformation of ${T^{q}_{(k)}}$ under commutation with ${S^{\mu}}$ is identical to that of an angular momentum eigenstate ${\ket{j=k,m=q}}$ under the action of ${S^{\mu}}$. Specifically, in Eq.~\eqref{eq:STalgebra}, commutators with ${S^{\pm}}$ act like ladder operators for ${T^q_{(k)}}$ of the same rank, and commuting ${T^q_{(k)}}$ with ${S^z}$ yields an eigenvalue ${q}$. The ${k=0}$ operators are scalars that commute with every component of ${S^{\mu}}$ and transform like ${j=0}$ singlet states. The ${k=1}$ vector operators in the spherical basis have the representation 
\begin{equation}\label{eq:opstatevec}
     T_{(1)}^{0}=T^{z},\quad T_{(1)}^{\pm 1}=\mp\frac{\big(T^{x}\pm iT^{y}\big)}{\sqrt{2}}
\end{equation}
and transform like ${j=1}$ triplet states.\par

Henceforth, we discuss irreducible spherical tensors and may simply refer to them as tensors for brevity, unless otherwise specified.\par 

\subsection{Hamiltonians and constraints}\label{sec:hamcon}
Hamiltonians in a translationally-invariant spin-${s}$ systems can be expressed in terms of spherical tensor operators. Much like Cartesian tensors in Eq.~\eqref{eq:reducible}, a ${K}$-local term in the Pauli basis can be decomposed into a sum of spherical tensors, i.e.
\begin{equation}
    S^{\mu_{i+1}}_{i+1}S^{\mu_{i+2}}_{i+2}\cdots S^{\mu_{i+K}}_{i+K}=\sum_{k=0}^{K}\sum_{\phi_k}\sum_{q=-k}^{k}a^{\phi_k}_{q}T^q_{(k),\phi_k,i},
\end{equation}
where ${i,i+1,\cdots i+K }$ are site indices and ${\phi_k}$ label the multiple representations of tensors of a given rank for ${k<K}$. Appendix~\ref{sec:completebasis} outlines such a procedure for a spin-1/2 chain. Since a translationally-invariant sum of spherical tensors
\begin{equation}\label{eq:sumtensor}
    T^{q}_{(k),\phi_k}=\sum_{i=0}^{L-1} T^{q}_{(k),\phi_k,i}
\end{equation}
is itself a spherical tensor with the same ${q,k}$  and ${\phi_k}$, we may write
\begin{equation}\label{eq:SumSpher}
\begin{aligned}        H&=H_{\text{sym}}+H_{\text{SG}}+H_{\text{A}}\\
        &=\sum_{k=0}^{K}\sum_{\phi_k}\sum_{q=-k}^{k}c_q^{\phi_k}T_{(k),\phi_k}^{q}\\
\end{aligned}
\end{equation}
where ${c_{q}^{\phi_k}}$ are complex coefficients. We emphasize that ${T_{(k),\phi_k}^{q}}$ are ${K}$-local and that the number of these operators is independent of ${L}$ for a fixed ${K}$. Since  ${(T^{q}_{(k)})^{\dagger}=(-1)^qT^{-q}_{(k)}}$,  the Hermiticity of ${H}$ implies ${(c^{\phi_k}_{q})^*=(-1)^qc^{\phi_k}_{-q}}$. In the symmetry-based framework of Sec.~\ref{sec:review}, ${H_{\text{sym}}}$ is a linear combination of the ${k=0}$ scalars, ${H_{\text{SG}}=\Omega S^z}$ is a ${k=1}$ vector, and the rest of the ${k>0}$ tensors can appear in ${H_{\text{A}}}$.\par

We now consider the constraints on ${\{c^{\phi_k}_{q}\}}$, which are imposed by the condition ${H_{\text{A}}\ket{\psi_n}=0}$ for all ${n\geq0}$. Note that it can be expanded as

\begin{equation}\label{eq:comexpansion}
    \begin{aligned}
    H_{\text{A}}\ket{\psi_n}&\equiv H_{\text{A}}\big(S^-\big)^n\ket{\psi_0}\\
    &=\sum_{r=0}^{n}\frac{n!}{(n-r)!r!} \big(S^-\big)^{n-r}h_r\ket{\psi_0},
    \end{aligned}
\end{equation}
where ${\{h_n\}}$ is the set of iterated commutators defined by
\begin{equation}\label{eq:iteratedcom}
\begin{aligned}
    h_0&\equiv H_{\text{A}},\quad h_{n+1}\equiv\big[h_{n},S^-\big]\quad \forall n\geq0.
\end{aligned}
\end{equation}
By considering the expansions with successively larger ${n\geq0}$, we see that the constraints reduce to
\begin{equation}\label{eq:commuteconstraint}
    h_n\ket{\psi_0}=0\quad\forall n\geq0.
\end{equation}
In terms of the tensors, they explicitly read
\begin{equation}\label{eq:explicitconstraint}
    \sum_{k=0}^{K}\sum_{\phi_k}\sum_{q=-k+n}^{k}c_{q}^{\phi_k}l^{-}(k,q,n)T_{(k),\phi_k}^{q-n}\ket{\psi_0}=0
\end{equation}
where 
\begin{equation}
    l^{-}(k,q,n)=\bigg(\frac{(k+q)!(k-q+n)!}{(k-q)!(k+q-n)!}\bigg)^{1/2}.
\end{equation}

If ${H_{\text{A}}}$ contains tensors with ${q\leq q^*}$  and the tensor with $q^*$ has rank ${k^*}$,
\begin{equation}\label{eq:nestedcom}
    h_{n}=0 \quad \forall n\geq (q^*+k^*)+1.
\end{equation}
The maximal rank of a one-body tensor is ${k=2s}$. Thus a ${K}$-body tensor has maximal rank ${2sK}$ which yields ${4sK+1}$ equations from the iterated commutators. Hence, for a spin-${s}$ Hamiltonian, the maximum number of nontrivial constraints is ${4sK+1}$. We note that Eq.~\eqref{eq:commuteconstraint} and Eq.~\eqref{eq:nestedcom} are equivalent to the RSGA conditions (iii), (iv) in Eq.~\eqref{eq:rsgacon}, where the order of the RSGA is identified as ${M=k^*+q^*}$, with ${M\leq4sK}$ as the upper bound.\par 
\par

\subsection{Advantages of the basis}\label{sec:utility}

\begin{table*}[tb]
  \renewcommand*{\arraystretch}{1}
    \begin{tabular}{ P{6cm}|P{6cm}|P{5cm} }
 \hline
 \hline
 Operator basis (and base state) &  Max. \# of equations & Max. \# of coefficients per equation  \\
 \hline
Pauli  & ${4sK+1=2K+1}$ & ${\sum_{N=1}^{K}\binom{K-1}{N-1}3^N=3\cdot 4^{K-1}}$\\
Spherical tensor (any ${\ket{\psi_0}}$) & ${2K+1}$ & ${3\cdot 4^{K-1}}$\\
Spherical tensor (${\ket{\psi_0}}$ with fixed ${z}$-magnetization) & ${\sum_{n=0}^{2K}(n+1)=(2K+1)(K+1)}$  & $\lesssim\sum_{N=1}^{K}\binom{K-1}{N-1}3^{N}{N}^{-1/2}$ \\
Spherical tensor (${\ket{\psi_0}}$ with highest weight and fixed ${z}$-magnetization) & $\sum_{k=0}^{K}(k^2+1)=(2K^2+K+6)(K+1)/6$  & $\sum_{N=1}^{K}\binom{K-1}{N-1}d_{N,k_{\text{max}}}$ \\
 \hline 
 \hline
    \end{tabular}
    \caption{A comparison of (the bounds of) the total number of equations that determine $H_A$ and the number of coefficients that enter each of these equations for different choices of operator basis and base states. Consider a 1d spin $s=1/2$ chain and up to ${K}$-local and range-${(K-1)}$ operators. Iterated commutators of ${H_{\text{A}}}$ with ${S^-}$ yield up to ${2K+1}$ equations in any basis (Eq.~\eqref{eq:commuteconstraint}), each involving up to ${3\cdot 4^{K-1}}$ coefficients. This applies to both the Pauli basis and the spherical tensor basis for an arbitrary $\ket{\psi_0}$. For a base state that has fixed ${z}$-magnetization, equations for different ${q}$ decouple, leading to Eq.~\eqref{eq:explicitconstraintwithmag}. The ${q=0}$ equation contains the most terms, but is still algebraically smaller as ${N}$ becomes large (Eq.~\eqref{eq:sumknumber}). If ${\ket{\psi_0}}$ is a highest weight state with fixed ${z}$-magnetization, we can further separate coefficients of different ${k}$, leading to Eq.~\eqref{eq:kqdecouple}. In the table, ${k_{\text{max}}}$ is the rank of the representation with maximal multiplicity for a given ${N}$ and ${K}$. }
    \label{fig:dimension}
\end{table*}


There are three main advantages to using spherical tensors to solve the constraint equations in Eq.~\eqref{eq:explicitconstraint}. We discuss how each of them simplifies the equations and summarize them in Table~\ref{fig:dimension}. \par 

First, the coefficients ${c^{\phi_k}_{q}}$ are easy to analytically compute for the largest values of $k, |q|$. Higher ${n}$ equations only involve ${c^{\phi_k}_{q}}$ with larger $q$, hence smaller number of coefficients. The equations become trivial after a maximum $n$ given by Eq.~\eqref{eq:nestedcom}. For instance, if we consider only tensors up to ${k=1}$ in the expansion of ${H_{\text{A}}}$, the ${n}$ we need to consider can be as small as ${n=2}$. In contrast, the coefficient of an operator ${O_a}$ in a reducible basis generically appears in every equation up to ${n=4sK}$.

Second, in solving Eq.~\eqref{eq:explicitconstraint} for $H_A$, the coefficients of the scalar ${(k=0)}$ operators can be disregarded. The co-efficients ${c^{\phi_0}_{0}}$ determine $H_\mathrm{sym} = \sum_{\phi_0} c_0^{\phi_0} T^0_{(0), \phi_0} $ and are only constrained by the requirement that $\ket{\psi_0}$ is an eigenstate of $H_\mathrm{sym}$. The number of coefficients that can be disregarded this way is given by 
\begin{equation}
 \sum_{N=1}^{K}\binom{K-1}{N-1}d_{N,0}\quad,
\end{equation}
where
\begin{equation}\label{eq:dimrepexact}
    d_{N,k}=\sum_{m=k}^{N}\frac{(-1)^{m+N}N!(2m)!(2k+1)}{m!(N-m)!(m-k)!(m+k+1)!}
\end{equation}
is the multiplicity of spin-${k}$ representations that are formed from the ${N}$-th tensor power of spin-{1} representations, or in our context, the number of rank ${k}$ tensor operators from the Kronecker product of ${N}$ one-body spin operators~\cite{polychronakos_composition_2016}. At large ${N}$, 
\begin{equation}\label{eq:dimredasym}
\begin{aligned}
    d_{N,k}\sim 3^N &N^{-\frac{3}{2}}\bigg(k+\frac{1}{2}\bigg)\exp{-\frac{3(k+\frac{1}{2})^2}{4N}}.
\end{aligned}
\end{equation}
\par

Third, for certain choices of ${\ket{\psi_0}}$, Eq.~\eqref{eq:explicitconstraint} is zero for each ${q}$ and ${k}$. This increases the total number of constraint equations~\footnote{Note that in some cases, some of these equations are linearly dependent. For example, in Appendix~\ref{sec:derivesol}, since ${T^q_{(1),\phi_1}\ket{\psi_0}}$ yield states with the same relative phase for all ${\phi_1}$ in the expansion, the real and imaginary parts of the linear equations decouple, making the ${n=2}$ equations redundant.}, but each of them involve far fewer coefficients, making an analytical solution more feasible. \par

When the base state ${\ket{\psi_0}}$ has fixed $z$-magnetization, each term in the $q$-sum in Eq.~\eqref{eq:explicitconstraint} is individually zero. That is,
\begin{equation}\label{eq:explicitconstraintwithmag}
    \sum_{k=0}^{K}l^{-}(k,q,n)\sum_{\phi_k} c_{q}^{\phi_k}T_{(k),\phi_k}^{q-n}\ket{\psi_0}=0.
\end{equation}
Consider an angular momentum eigenstate ${\ket{j,m}}$ with ${m=-j,\cdots,j}$. Using Eq.~\eqref{eq:STalgebra}, one can show that
 \begin{equation}\label{eq:wetheorem}
    \bra{j',m'}T^q_{(k)}\ket{j,m}\propto\bra{j',m'}\ket{j,m;k,q}\propto \delta_{m', m+q}.
\end{equation}
Eq.~\eqref{eq:explicitconstraintwithmag} follows as ${T^{q_1}_{(k_1),\phi_{k_1}}\ket{\psi_0}}$ and ${T^{q_2}_{(k_2),\phi_{k_2}}\ket{\psi_0}}$ are linearly independent for any ${\phi_{k_1}}$ and ${\phi_{k_2}}$ if ${q_1\neq q_2}$. The ${q=0}$, ${n=0}$ equation contains the most number of coefficients. At large $N$, this is given by 
\begin{equation}\label{eq:sumknumber}
    \sum_{k=0}^{N}d_{N,k}\sim 3^{N}N^{-\frac{1}{2}},
\end{equation}
which is smaller than ${3^N}$, the total number of ${N}$-body operators.

When the base state ${\ket{\psi_0}}$ is a highest-weight state with fixed $z$-magnetization (e.g. $\ket{\psi_0}=\ket{\psi_{0,0}}$ in Eq.~\eqref{eq:multimagnonstate}), each term in the $k$-sum in Eq.~\eqref{eq:explicitconstraintwithmag} is also individually zero. Consider ${n=2K}$, so that the only non-trivial equation is
\begin{equation}\label{eq:maxKconstraint}
    l^-(K,K,2K)\sum_{\phi_K}c^{\phi_K}_{K}T^{-K}_{(K),\phi_K}\ket{\psi_0}=0.
\end{equation}
Since ${l^-(K,K,2K)\neq0}$, the sum must vanish. Setting ${n=2K-1}$, we similarly have
\begin{equation}\label{eq:nextmaxKconstraint}
    l^-(K,K-1,2K-1)\sum_{\phi_K}c^{\phi_K}_{K-1}T^{-K}_{(K),\phi_K}\ket{\psi_0}=0,
\end{equation}
\begin{equation}
    l^-(K,K,2K-1)\sum_{\phi_K}c^{\phi_K}_{K}T^{-K+1}_{(K),\phi_K}\ket{\psi_0}=0.
\end{equation}
Using Eq.~\eqref{eq:STalgebra} and the fact that ${S^+}$ annihilates highest-weight states, we can apply ${\big(S^+\big)^M}$ to the left of Eq.~\eqref{eq:maxKconstraint} and show that
\begin{equation}\label{eq:highweighttrick}
\begin{aligned}
0&= \sum_{\phi_K}c^{\phi_K}_{K}\big(S^+\big)^{{M-1}}S^+T^{-K}_{(K),\phi_K}\ket{\psi_0}\\
&\propto\sum_{\phi_K}c^{\phi_K}_{K}\big(S^+\big)^{{M-1}}T^{-K+1}_{(K),\phi_K}\ket{\psi_0}\propto\sum_{\phi_K}c^{\phi_K}_{K}T^{-K+M}_{(K),\phi_K}\ket{\psi_0}\\
&\implies\sum_{\phi_K}c^{\phi_K}_{K}T^{q}_{(K),\phi_K}\ket{\psi_0}=0\\
\end{aligned}
\end{equation}
for all ${q\in\{-K,\cdots,K\}}$. Likewise, using Eq.~\eqref{eq:nextmaxKconstraint},
\begin{equation}
    \sum_{\phi_K}c^{\phi_K}_{K-1}T^{q}_{(K),\phi_K}\ket{\psi_0}=0\quad\forall q\in\{-K,\cdots,K\}.
\end{equation}
Setting ${n=2(K-1)}$, we may use the above to deduce that

\begin{equation}\label{eq:nextnextmaxKconstraint}
    l^-(K-1,K-1,2K-2)\sum_{\phi_{K-1}}c^{\phi_{K-1}}_{K-1}T^{-(K-1)}_{(K-1),\phi_{K-1}}\ket{\psi_0}=0,
\end{equation}
\begin{equation}
  l^-(K,K-2,2K-2)\sum_{\phi_K}c^{\phi_K}_{K-2}T^{-K}_{(K),\phi_K}\ket{\psi_0}=0.
\end{equation}
Repeating the line of argument from Eq.~\eqref{eq:maxKconstraint} to Eq.~\eqref{eq:nextnextmaxKconstraint} for progressively smaller ${n}$ and ${q}$, we arrive at
\begin{equation}\label{eq:kqdecouple}
    \sum_{\phi_k}c_{q_a}^{\phi_k}T^{q_b}_{(k),\phi_{k}}\ket{\psi_0}=0
\end{equation}
for any ${k}$ and ${|q_a|,|q_b|\leq k}$. Thus, each term in the ${k}$-sum in Eq.~\eqref{eq:explicitconstraintwithmag} is individually zero. This implies the constraints may be solved for any fixed ${k}$ and ${q}$.

\section{Embedding magnon scars}\label{sec:magnonscar}
In this section, we present a family of periodic 1d scarred models constructed from the spherical tensor formalism. These models host scarred states with any number of ${p=0}$ magnons and one (or any number of) ${p_0\neq0}$ magnon(s). 

\subsection{Target states}\label{sec:targetstates}
The magnon states are the spin flip excitations of the fully polarized state
\begin{equation}\label{eq:fullpol}
    \ket{\boldsymbol{s}}\equiv\bigotimes_{i=0}^{L-1}\ket{s}_i.
\end{equation}
For ${s=1/2}$, which we shall discuss primarily, we denote ${\ket{\boldsymbol{s}}=\bigotimes_{i=0}^{L-1}\ket{\uparrow}_i=\ket{\text{up}}}$. Likewise, ${\bigotimes_{i=0}^{L-1}\ket{\downarrow}_i=\ket{\text{down}}}$. The creation operator of a magnon with momentum ${p=2\pi m/L}$ atop $\ket{\boldsymbol{s}}$ is given by

\begin{equation}\label{eq:magnonop}
    Q^{-}(p)=\sum_{i=0}^{L-1}e^{- ipr_i}S^{-}_{i},\quad Q^{-}(p=0)\equiv S^{-},
\end{equation}
where ${r_i}$ is a site label, and ${Q^{+}(p)=\big(Q^-(p)\big)^{\dagger}}$. The target states are parametrized by ${n}$, the number of ${p=0}$ magnons and ${N}$, the number of ${p_0\neq0}$ magnons
\begin{equation}\label{eq:multimagnon}
\begin{aligned}
    &\ket{\psi_{n,N}(p_0)}=\big(S^-\big)^n\big(Q^-(p_0\neq0)\big)^N\ket{\text{up}},\\
    &\ket{\psi_{0,0}(p_0)}\equiv\ket{\text{up}},\\
    n&=\begin{cases}
    0,1,\cdots,L,\quad N=0\\
    0,1,\cdots,L-N-1,\quad 1\leq N\leq L-1\\
    \end{cases}.
\end{aligned}
\end{equation}
We regard ${\{\ket{\psi_{0,N}(p_0)}\}}$ as the set of base states, and ${S^-}$ as the ladder operator that generates the rest of the multiplets. 
\par 

The target states are generally not eigenstates of the total spin operator ${S^2}$. The exceptions are ${\ket{\psi_{n,0}(0)}}$ and ${\ket{\psi_{n,1}(p_0)}}$, which span the ${j=Ls}$ and ${j=Ls-1}$ multiplets. In addition, ${\ket{\psi_{0,L-1}(p_0)}}$ are also eigenstates with ${j=Ls-1}$. This is because we can write
\begin{equation}
\begin{aligned}
    &\big(Q^-(p)\big)^{L}\ket{\text{up}}\propto\ket{\text{down}}\\
    \implies & \big(Q^-(p)\big)^{L-1}\ket{\text{up}}\propto Q^+(p)\ket{\text{down}}
\end{aligned}
\end{equation}
which is an eigenstate of ${S^2}$.\par 

Note that ${Q^-(p_0)}$ can also be regarded as the ladder operator for the set of base states ${\{\ket{\psi_{n,0}(p_0)}\}}$ within the symmetry-based framework discussed in Sec.~\ref{sec:review}. This is because the set of operators
\begin{equation}
   Q^+(p),Q^-(p),Q^z(p)=\frac{\big[Q^+(p),Q^-(p)\big]}{2}=S^z
\end{equation}
is associated with the Lie algebra of a ${Q}$-SU(2) symmetry with the commutation relations ${\big[S^z,Q^{\pm}(p)\big]=\pm Q^{\pm}(p)}$ and a corresponding Casimir operator ${Q^2(p)=\{Q^+(p),Q^-(p)\}/2+(S^z)^2}$.\par 

In the rest of the main text, we exclusively consider models on a ${L}$-site periodic ${s=1/2}$ chain.

\subsection{Shiraishi-Mori annihilators}\label{sec:SManni}
The spherical tensor formalism yields a complete basis of translationally-invariant operators which includes the subset of Shiraishi-Mori (SM) annihilators. These annihilators are constructed from local configurations orthogonal to those spanned by the target states. Note that such operators need not be local projectors as introduced in Ref.~\cite{shiraishi_systematic_2017}.\par 

The minimum number of sites with local configurations orthogonal to those spanned by the target states is three. Consider three contiguous sites $i$, $i+1$ and $i+2$. It can be shown that
\begin{equation}
\begin{aligned}
    \ket{v_a(p_0)}_i&=\frac{1}{2\sqrt{2-\cos{p_0}-\cos^2{p_0}}}\big[({1-e^{ip_0}})\ket{\uparrow\uparrow\downarrow}_i\\
    &+({e^{ip_0}-1})\ket{\uparrow\downarrow\uparrow}_i+e^{ip_0}({1-e^{ip_0}})\ket{\downarrow\uparrow\uparrow}_i\big]\\
    \ket{v_b(p_0)}_i&=\frac{1}{2\sqrt{2-\cos{p_0}-\cos^2{p_0}}}\big[({1-e^{ip_0}})\ket{\downarrow\downarrow\uparrow}_i\\
    &+({e^{ip_0}-1})\ket{\downarrow\uparrow\downarrow}_i+e^{ip_0}({1-e^{ip_0}})\ket{\uparrow\downarrow\downarrow}_i\big]
\end{aligned}
\end{equation}
are orthogonal to the local configurations that appear in ${\{\ket{\psi_{n,N=0,1}(p_0)}\}}$. Thus, we can construct SM annihilators from
\begin{equation}\label{eq:SMproj}
\begin{aligned}
    &P^{\text{SM}}_{1,i}(p_0)=\ket{v_a(p_0)}_i\bra{v_a(p_0)}_i,\\
    &P^{\text{SM}}_{2,i}(p_0)=\ket{v_b(p_0)}_i\bra{v_b(p_0)}_i,\\
    &P^{\text{SM}}_{3,i}(p_0)=\frac{\ket{v_a(p_0)}_{i}\bra{v_b(p_0)}_{i}+\ket{v_b(p_0)}_{i}\bra{v_a(p_0)}_{i}}{2},\\
    &P^{\text{SM}}_{4,i}(p_0)=\frac{\ket{v_a(p_0)}_{i}\bra{v_b(p_0)}_{i}-\ket{v_b(p_0)}_{i}\bra{v_a(p_0)}_{i}}{2i}.
\end{aligned}
\end{equation}

Although the operators in Eq.~\eqref{eq:SMproj} are only designed to locally annihilate ${\ket{\psi_{n,N=0,1}(p_0)}}$ states, they may annihilate other multi-magnon eigenstates. First, when ${p_0\neq\pi}$, ${P^{\text{SM}}_{1,i}(p_0)}$ annihilates states with any number of $p=0$ and $p=p_0$ magnons. That is, 
\begin{equation}\label{eq:allmagnon}
    P^{\text{SM}}_{1,i}(p_0)\ket{\psi_{n,N}(p_0)}=0\quad\forall n,N\geq0,
\end{equation}
The Hamiltonian $H_{\text{A}}=\sum_i P^{\text{SM}}_{1,i}(p_0)$ thus provides an example of a \emph{multi-$p_0$-magnon model}, see Sec.~\ref{sec:generalHam}. We confirm in Sec.~\ref{sec:multipnumerics} that the Hamiltonian is otherwise thermalizing (within each $S^z$ sector as $S^z$ is a good quantum number).

When ${p_0=\pi}$, Eq.~\eqref{eq:allmagnon} applies to all four operators ${\{P^{\text{SM}}_{a,i}(\pi)\}_{a=1}^{4}}$. We thus can also construct \emph{multi-$\pi$-magnon models} of the SM type that are otherwise thermalizing.  

A comment is in order. If the chain is periodic, translationally invariant and conserves the total $z$-magnetization, then all single-magnon states are eigenstates. As $P^{\text{SM}}_{a,i}(p_0)$ commutes with $S^z$ for $a=1,2$, the multi-$p_0$-magnon model described above has all single-magnon states as eigenstates. 

To determine ${H_{\text{A}}}$ that are beyond the SM formalism, we project out the annihilators in Eq.~\eqref{eq:SMproj} in Sec.~\ref{sec:stbasis}. 

\subsection{Spherical tensors in scarred Hamiltonians}\label{sec:stbasis}


 We find that ${H_{\text{sym}}}$ and ${H_{\text{A}}}$ can be constructed from the following representations of spherical tensors. By considering the fully polarized base state ${\ket{\psi_0}=\ket{\text{up}}}$, we show in Appendix~\ref{sec:completebasis} that it is relatively easy to construct these terms analytically. \par 

${H_{\text{sym}}}$ is spanned by four scalars. Aside from the identity, there are the range-${R}$ Heisenberg (denoted by H-${R}$) terms
\begin{equation}\label{eq:heisenberg}
    H_{R}^{\text{H}}=\sum_{i=0}^{L-1}\boldsymbol{S}_{i}\cdot\boldsymbol{S}_{i+R},
\end{equation}
We only use the nearest-neighbor (${R=1}$) and next-nearest-neighbor (${R=2}$) terms. The fourth is the `scalar triple product' (ST) term
\begin{equation}\label{eq:scalarproduct}
    H^{\text{ST}}=\epsilon^{\mu\nu\lambda}\sum_{i=0}^{L-1}S^{\mu}_{i-1}S^{\nu}_{i}S^{\lambda}_{i+1}.
\end{equation}
\par
${H_\text{A}}$ can be constructed from four sets of vectors and one set of rank-2 tensor operators, which are orthogonal to the 3-local SM annihilators in Eq.~\eqref{eq:SMproj}. Two of these are the range-${R}$ Dzyaloshinskii-Moriya interaction (denoted by DM-${R}$) terms
\begin{equation}\label{eq:DMR}
    H_{R}^{\text{DM}}(\mu)=\epsilon^{\mu\nu\lambda}\sum_{i=0}^{L-1}S^{\nu}_{i}S_{i+R}^{\lambda},\quad \mu=x,y,z
\end{equation}
with ${R=1,2}$. Then, there are two sets of vectors, namely, the `vector triple product plus' (VTP) terms
\begin{equation}\label{eq:VTP}
\begin{aligned}
    H^{\text{VTP}}(\mu)=\sum_{i=0}^{L-1}&{S}_{i-1}^{\mu}(\boldsymbol{S}_{i}\cdot\boldsymbol{S}_{i+1})-2{S}^{\mu}_{i}(\boldsymbol{S}_{i-1}\cdot\boldsymbol{S}_{i+1})\\
    &+(\boldsymbol{S}_{i-1}\cdot\boldsymbol{S}_{i}){S}^{\mu}_{i+1},
\end{aligned}
\end{equation}
and the `vector triple product minus' (VTM) terms
\begin{equation}\label{eq:VTM}
    H_{R}^{\text{VTM}}(\mu)=\sum_{i=0}^{L-1}{S}_{i-1}^{\mu}(\boldsymbol{S}_{i}\cdot\boldsymbol{S}_{i+1})-(\boldsymbol{S}_{i-1}\cdot\boldsymbol{S}_{i}){S}^{\mu}_{i+1},
\end{equation}
which are so named because each local term can be expressed in the following combinations in Eq.~\eqref{eq:VTP} and Eq.~\eqref{eq:VTM}:
\begin{equation}
    \begin{aligned}
    &\boldsymbol{S}_{k}\times(\boldsymbol{S}_{i}\times\boldsymbol{S}_j)+\boldsymbol{S}_{i}\times(\boldsymbol{S}_{k}\times\boldsymbol{S}_j)\\
    &=\boldsymbol{S}_{i}(\boldsymbol{S}_{j}\cdot \boldsymbol{S}_{k})-2\boldsymbol{S}_{j}(\boldsymbol{S}_{i}\cdot \boldsymbol{S}_{k})+(\boldsymbol{S}_{i}\cdot \boldsymbol{S}_{j})\boldsymbol{S}_{k},\\    &\boldsymbol{S}_{k}\times(\boldsymbol{S}_{i}\times\boldsymbol{S}_j)-\boldsymbol{S}_{i}\times(\boldsymbol{S}_{k}\times\boldsymbol{S}_j)\\
    &=\boldsymbol{S}_{i}(\boldsymbol{S}_{j}\cdot \boldsymbol{S}_{k})-(\boldsymbol{S}_{i}\cdot \boldsymbol{S}_{j})\boldsymbol{S}_{k}.
    \end{aligned}
\end{equation}
%
%
Finally, there are five terms (denoted by RT or Rank-Two) terms that can be expressed as linear combinations of a set of rank-2 tensor operators. Namely (explicitly given in Appendix~\ref{sec:commutatorsbase}),
\begin{equation}\label{eq:RTTlist}
\begin{aligned}
H^{\text{RT}}(1)=\frac{T^{1}_{(2),a}+T^{-1}_{(2),a}}{4},&\quad H^{\text{RT}}(2)=\frac{T^{-1}_{(2),a}-T^{1}_{(2),a}}{4i}\\
    H^{\text{RT}}(4)&=-i{T^{0}_{(2),a}}/2,\\
H^{\text{RT}}(3)=\frac{T^{-2}_{(2),a}-T^{2}_{(2),a}}{2},&\quad H^{\text{RT}}(5)=\frac{i(T^{2}_{(2),a}+T^{-2}_{(2),a})}{2}\\
\end{aligned}
\end{equation}
where up to normalizations,
\begin{equation}
\begin{aligned}
    {T^{0}_{(2),a}}=\sum_{i=0}^{L-1}&S^z_{i-1}(S^-_{i}S^+_{i+1}-S^+_{i}S^-_{i+1})\\
    &+(S^-_{i-1}S^+_{i}-S^+_{i-1}S^-_{i})S^z_{i+1}\\
    &+2S^z_{i}(S^-_{i-1}S^+_{i+1}-S^+_{i-1}S^-_{i+1})\\
\end{aligned}
\end{equation}
\begin{equation}
\begin{aligned}
    {T^{\pm1}_{(2),a}}=\sum_{i=0}^{L-1}&(S^{\pm}_{i-1}S^{\mp}_{i+1}-S^{\mp}_{i-1}S^{\pm}_{i+1})S^{\pm}_{i}\\
    &+2(S^z_{i-1}S^{\pm}_{i+1}-S^{\pm}_{i-1}S^z_{i+1})S^z_{i}
\end{aligned}
\end{equation}
\begin{equation}
    \begin{aligned}
    T^{\pm2}_{(2),a}=\pm\sum_{i=0}^{L-1}S^{\pm}_{i}(S^{\pm}_{i-1}S^z_{i+1}-S^z_{i-1}S^{\pm}_{i+1}).
    \end{aligned}
\end{equation}
Aside from the VTP, all of the operators in ${H_{\text{A}}}$ are anti-symmetric with respect to spatial inversion. This symmetry is a useful guiding principle to construct representations of tensor operators that annihilate the fully polarized base state because of the cancellation of terms with their spatially-inverted counterparts.\par

\subsection{General forms of scarred models}\label{sec:generalHam}
Using the operators in Secs.~\ref{sec:SManni} and ~\ref{sec:stbasis}, we now write down families of Hamiltonians that host different sets of magnon scars. In all these cases, we can write ${H_{\text{SG}}=\Omega S^z}$ and
\begin{equation}
    H_{\text{sym}}=J_0+J_1 H^{\text{H}}_1+J_2 H^{\text{H}}_2 +J_3 H^{\text{ST}}
\end{equation}
where ${\{J_i\}_0^3}$ are free parameters. The form of ${H_{\text{A}}}$ depends on the target states to be embedded as scars. We separate out the SM terms $H^{\text{SM}}_{\text{A}}(p_0)$ that locally annihilate the target states and write:
\begin{equation}
    H_{\text{A}} = H'_{\text{A}}(p_0) + H^{\text{SM}}_{\text{A}}(p_0).
\end{equation}
The SM annihilators in Eq.~\eqref{eq:SMproj} determine $H^{\text{SM}}_{\text{A}}(p_0)$:
\begin{align}
   H^{\text{SM}}_{\text{A}}(p_0)= \sum_{a} K_a \sum_i P^{\text{SM}}_{a,i}(p_0)
\end{align}
where ${\{K_a\}}$ are free parameters in all but the multi-$p_0$ magnon model.
\par 
\paragraph{Zero-magnon models.} States with any number of ${p=0}$ magnons (the multiplet ${\{\ket{\psi_{n,0}(0)}\}}$) are annihilated by any linear combination of the operators from Eq.~\eqref{eq:DMR} to Eq.~\eqref{eq:RTTlist}, and by the SM annihilators in Eq.~\eqref{eq:SMproj} for any $p_0$.

\paragraph{Single-${p_0}$-magnon models.} States with any number of ${p=0}$ magnons and up to one ${p_0\neq0,\pi}$ magnon (${\{\psi_{n,N=0,1}(p_0)\}}$) are annihilated by ${H'_{\text{A}}(p_0)}$ of the form
\begin{equation}\label{eq:singlep}
    H'_{\text{A}}(p_0)=\sum_{\mu=x,y,z}J^{V}_{\mu}V^{\mu}(p_0,1/2)
\end{equation}
where ${\{J^{V}_{\mu}\}}$ are free parameters and
\begin{equation}\label{eq:vectorterm}
\begin{aligned}
    V^{\mu}(p_0,s)&=2s\sin(p_0)H^{\text{DM}}_{1}(\mu)+s\tan\bigg(\frac{p_0}{2}\bigg)H^{\text{DM}}_{2}(\mu)\\
    &+H^{\text{VTP}}(\mu).
\end{aligned}
\end{equation}
The explicit derivation of the vector term ${V^{\mu}(p_0,s)}$ is provided in Appendix \ref{sec:derivesol}.

\paragraph{Single-${\pi}$-magnon models.} States with any number of ${p=0}$ magnons and up to one ${p_0=\pi}$ magnon (${\{\psi_{n,N=0,1}(\pi)\}}$) are annihilated by 
\begin{equation}\label{eq:annihilatepi}
\begin{aligned}
    H'_{\text{A}}(\pi)&=\sum_{\mu=x,y,z}J_{\mu}^{\text{DM2}}H^{\text{DM}}_2(\mu)+\sum_{\lambda=1}^{5}J_{\lambda}^{\text{RT}}H^{\text{RT}}(\lambda)\\
\end{aligned}
\end{equation}
where ${\{J_{\mu}^{\text{DM2}}\}}$ and ${\{J_{\lambda}^{\text{RT}}\}}$ are free parameters. The form of ${H'_{\text{A}}(\pi)}$ is distinct from Eq.~\eqref{eq:vectorterm} because of the following. Let ${\phi_1}$ label the DM1, VTP, VTM representations of ${k=1}$ tensors. Of all the ${k>0}$ tensors in Sec.~\ref{sec:stbasis}, these are the only tensors that do not individually annihilate the target states. In particular, since ${T^{-1}_{(1),\phi_1}Q^-(\pi)\ket{\text{up}}\neq0}$, and the tensors are linearly independent, their coefficients ${c_{q}^{\phi_1}}$ are zero for all ${q}$. \par 
\paragraph{Multi-${\pi}$-magnon models.} 
In the absence of the nearest-neighbor Heisenberg term i.e. ${J_1=0}$, states with any number of ${p=0}$ and ${p_0=\pi}$ magnons ((${\{\psi_{n,N}(\pi)\}}$) are preserved as eigenstates of ${H}$ with the same ${H'_{\text{A}}}$ in Eq.~\eqref{eq:annihilatepi}. 

\paragraph{Multi-${p_0}$-magnon models.}\label{sec:multiparag}
The set of states with any number of ${p=0}$ and ${p_0\neq0,\pi}$ magnons ((${\{\psi_{n,N}(p_0)\}}$) cannot be annihilated by any 3-local non-SM Hamiltonian. That is, ${H'_\text{A}(p_0)=0}$. Furthermore, $K_{2,3,4}=0$ and ${J_{1,2,3}=0}$. The simplest model with a non-trivial ${H_{\text{sym}}}$ is thus given by
\begin{equation}\label{eq:multipHam}
    H_{\text{sym}}=J_M H^{\text{H}}_{M},\quad H_\text{A}=K_1\sum_{i=0}^{L-1}P^{\text{SM}}_{1,i}(p_0)
\end{equation}
where ${H^{\text{H}}_{M}}$ is the range-${M}$ Heisenberg term and ${M}$ is an integer multiple of the denominator of ${p_0/2\pi}$.

\section{Numerical simulations}\label{sec:simulation}
This section presents numerical results obtained by exactly diagonalizing three models from  Sec.~\ref{sec:generalHam}. Based on the level statistics and the entanglement entropy of the eigenstates, we conclude that the models are scarred by the targeted magnon states. In addition, we discuss the quench dynamics of initial states which can be used to probe the ${p_0\neq0}$ magnon scars.

\subsection{Single-${p_0}$-magnon model}\label{sec:2pi3magnon}

\begin{figure}
\subfigure{\includegraphics[width=\linewidth]{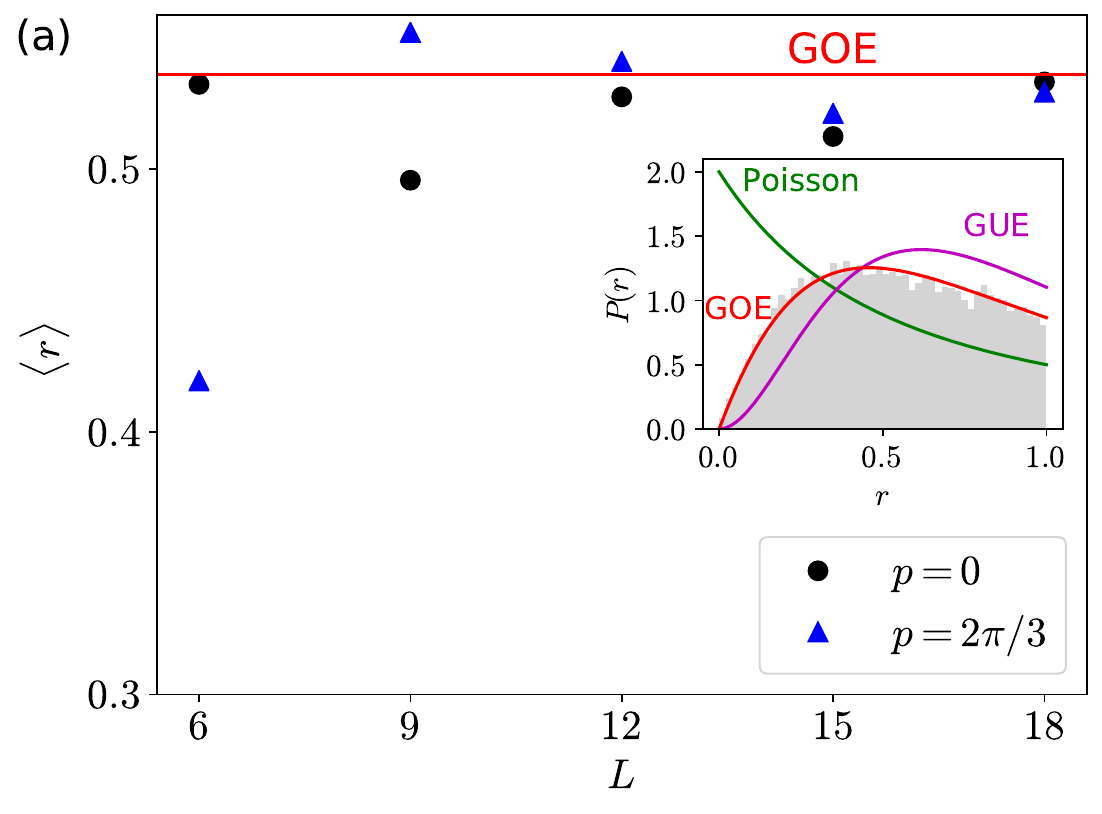}\label{fig:pr2pi3}}
    \subfigure{
			\includegraphics[width=\linewidth]{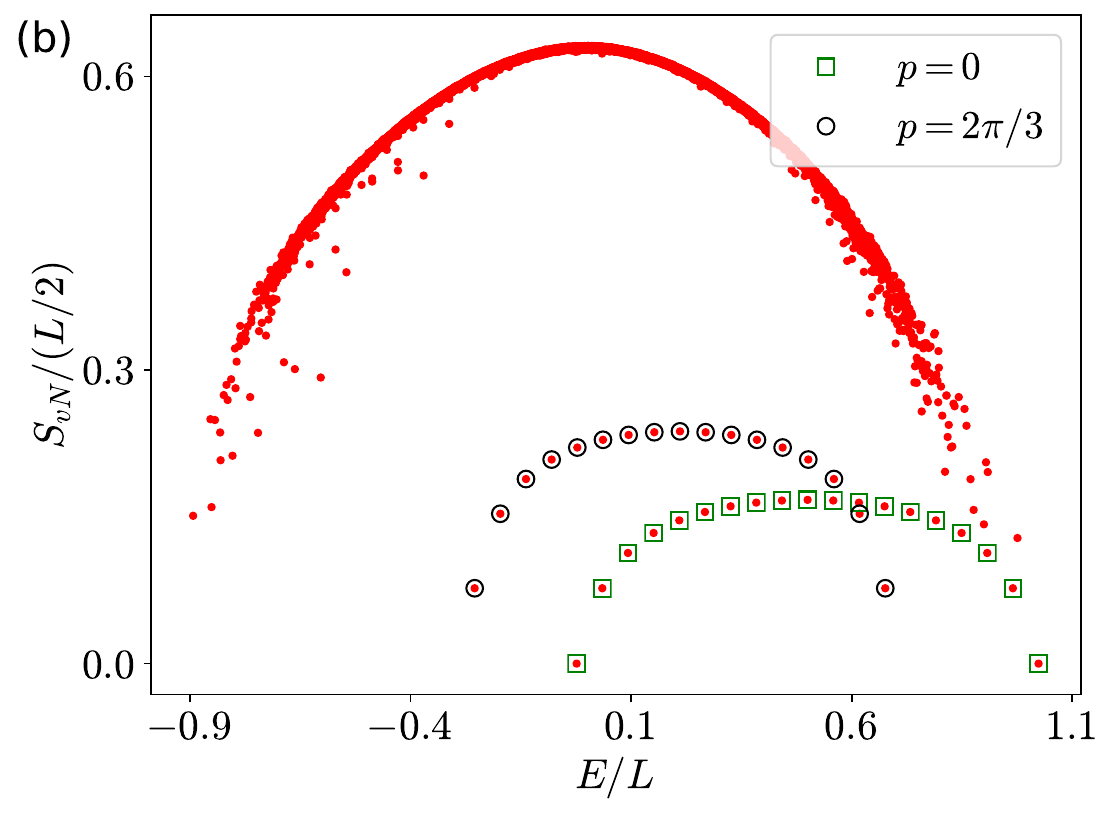}\label{fig:2pi3SE}
			}

		\caption{(\textbf{a}) Probes of the level statistics of ${H_{2\pi/3}}$ (parameters given in Eq. \eqref{eq:h2pi3parameters}) using the set of $r_n$, the level spacing ratios from \eqref{eq:rdef}. The main plot shows that the mean ${\langle r_n \rangle}$ values in the ${p=0}$ and ${p=2\pi/3}$ sectors both approach the GOE value ${\approx0.5359}$ \cite{atas_distribution_2013} with increasing system size ${L}$. The inset gives a normalized histogram of the $r_n$ (${p=2\pi/3}$ sector, ${L=18}$), $P(r)$, which closely matches the GOE prediction given in \cite{atas_distribution_2013}. (\textbf{b}) The half-chain entanglement entropy density plotted against the energy density of each eigenstate of ${H_{2\pi/3}}$ in the ${p=0}$ and ${p=2\pi/3}$ momentum sectors. At ${L=18}$, the 17 ${\{\psi_{n,1}(p_0)\}}$ scarred states are marked by circles, whereas the 19 ${\{\psi_{n,0}(p_0)\}}$ scarred states are marked by squares. Aside from these 36 states, the entropy density appears to be a smooth function of the energy density, which is one of the hallmarks of a thermalizing model.}
\label{fig:main1}

\end{figure}
We show that the model ${H_{2\pi/3}}$ with ${H_{\text{A}}=H'_{\text{A}}(p_0)}$ given by Eq.~\eqref{eq:singlep} with parameters
\begin{equation}\label{eq:h2pi3parameters}
    \begin{aligned}
    H_{2\pi/3}:
        (p_0,J_0,J_1,J_2,J_3,\Omega)&=(2\pi/3,0,1,1,1.7,\pi/3)\\
        (J^{\text{V}}_{x},J^{\text{V}}_{y},J^{\text{V}}_{z})&=(1,0,0)
    \end{aligned}
\end{equation}
is scarred by the states ${\{\psi_{n,N=0,1}(2\pi/3)\}}$. For simplicity, we do not include the SM annihilators.
\paragraph{Symmetries and level statistics.}
Aside from translational symmetry, the model also possess an anti-unitary symmetry. Let ${\mathcal{T}}$ be the time reversal operator, ${\mathcal{R}^{y}_{\pi}}$ be the operator which rotates all spins along the ${y}$ axis by ${\pi}$, and  ${\mathcal{I}}$ be the spatial inversion operator about a site/bond (for odd/even ${L}$). ${H_{2\pi/3}}$ is invariant under the operation 
\begin{equation}
    \mathcal{I}\mathcal{R}^{y}_{\pi}\mathcal{T}H_{2\pi/3}\mathcal{T}^{-1}(\mathcal{R}^{y}_{\pi})^{-1}\mathcal{I}^{-1}=H_{2\pi/3}.
\end{equation} 
To test whether the (sorted) spectrum ${\{E_n\}}$ is thermalizing, we examine the distribution of level spacing ratios \cite{atas_distribution_2013}, defined by
\begin{equation}\label{eq:rdef}
    r_n=\frac{\text{min}(s_n,s_{n-1})}{\text{max}(s_n,s_{n-1})}
\end{equation}
where ${s_n=E_{n+1}-E_{n}}$ are the consecutive energy level spacings. Fig. \ref{fig:pr2pi3} indicates that the level statistics of the model is that of a thermalizing model and because of the presence of the anti-unitary symmetry, is given by the Gaussian Orthogonal Ensemble (GOE). 

\paragraph{Entanglement entropy.}
To further test thermalization on an eigenstate level, we also consider the half-chain von Neumann entanglement entropy ${S_{\text{vN}}}$ of each eigenstate. In systems described by the ETH, the entanglement entropy density, ${S_{\text{vN}}/(L/2)}$, clusters around a smooth function of the energy. Fig. \ref{fig:2pi3SE} shows that this is indeed the case except for the outlying scarred states, which constitutes a weak violation of the ETH.\par 
\paragraph{Quench dynamics and revival.} Scarred states with one ${p_0\neq0}$ magnon can be probed by quenching from a family of initial states, given by the coherent state
\begin{equation}
\begin{aligned}
    \ket{\Phi}&=\mathcal{N}\exp{cS^-}Q^-(2\pi/3)\ket{\text{up}}\\
    &=\mathcal{N}\sum_{m=0}^{\infty}\frac{(cS^-)^{m}}{m!}Q^-(2\pi/3)\ket{\text{up}}
\end{aligned}
\end{equation}
where ${c}$ is a complex number and ${\mathcal{N}}$ is the normalization constant. The state evolves within the scarred manifold as
\begin{equation}
\begin{aligned}
    \ket{\Phi(t)}&=\mathcal{N}e^{-iH_{2\pi/3}t}\ket{\Phi}\\
    &=\mathcal{N}e^{-iE_0 t}\exp{ce^{i\Omega t}S^-}Q^-(2\pi/3)\ket{\text{up}}
\end{aligned}
\end{equation}
where ${E_0}$ is the energy of the base state ${Q^-(2\pi/3)\ket{\text{up}}}$. 
On the other hand, since ${(S^-_{i})^2=0}$ for spin-${1/2}$ and ${\exp{cS^-}=\prod_{i=0}^{L-1}\exp{cS^-_{i}}}$, we may write
\begin{equation}\label{eq:phiT}
\begin{aligned}
    \ket{\Phi(t)}&=\mathcal{N}e^{-iE_0t}Q^-(2\pi/3)\prod_{i=0}^{L-1}(1+ce^{i\Omega t}S^-_{i})\ket{\text{up}}\\
    &=\mathcal{N}e^{-iE_0t}Q^-(2\pi/3)\prod_{i=0}^{L-1}\bigg(\ket{\uparrow}_{i}+ce^{i\Omega t}\ket{\downarrow}_{i}\bigg).
\end{aligned}
\end{equation}
%
The entanglement entropy of ${\ket{\Phi(t)}}$ is finite but constant in time. For a half cut, this is given by ${S_{\text{vN}}=\log{2}}$. Using this expression, one can show that the fidelity shows perfect revival as
\begin{equation}
    \begin{aligned}
        \mathcal{F}(t)&=|\braket{\Phi}{\Phi(t)}|^2\\
        &=\bigg(\frac{1+|c|^4+2|c|^2\cos\Omega t}{1+|c|^4+2|c|^2}\bigg)^{L-2}.
    \end{aligned}
\end{equation}

\subsection{Multi-${\pi}$-magnon model}\label{sec:pipi}
\begin{figure}
\includegraphics[width=\linewidth]{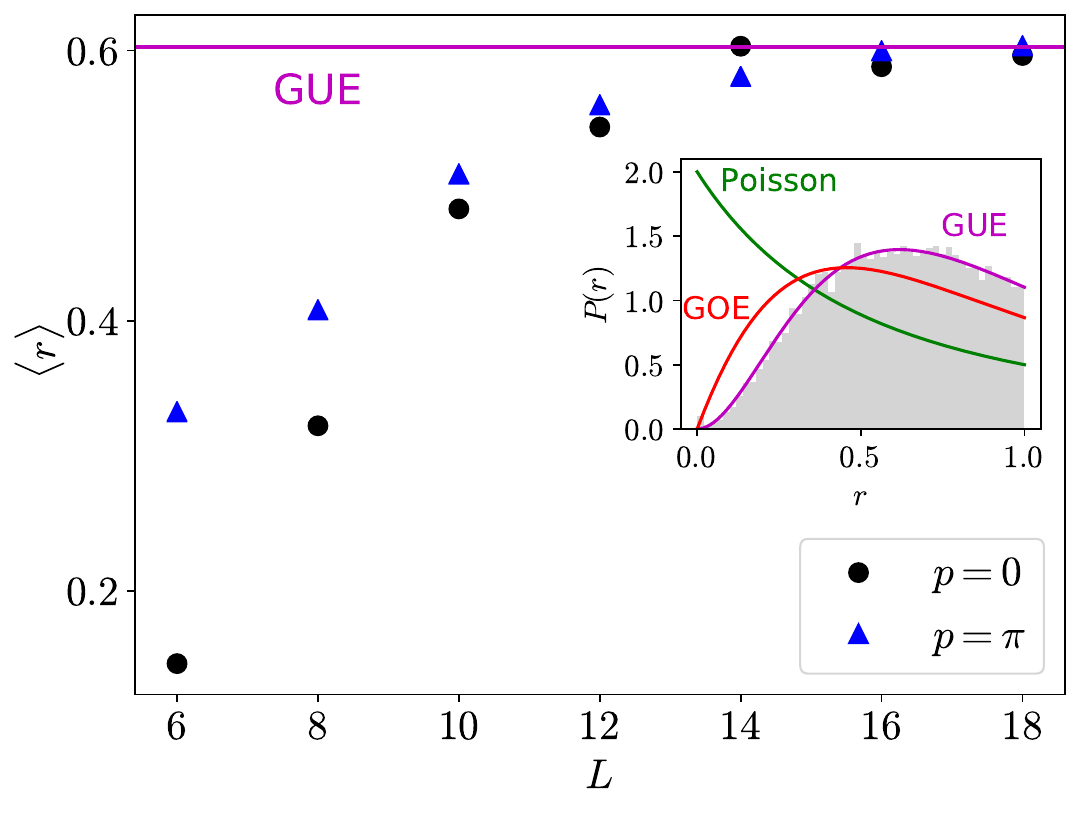}

		\caption{Probes of the level statistics of ${H_{\pi}}$ (parameters given in Eq. \eqref{eq:multipimagnon}) using the set of $r_n$, the level spacing ratios from \eqref{eq:rdef}. The main plot shows the mean $\langle r\rangle $ values in the ${p=0}$ and ${p=\pi}$ sectors asymptoting to the GUE value ${\approx 0.6027}$ \cite{atas_distribution_2013} with increasing system size ${L}$. The inset with a histogram of the $r_n$ (${p=\pi}$ sector, ${L=18}$), $P(r)$, closely matches the GUE prediction in \cite{atas_distribution_2013}. }
\label{fig:prpi}

\end{figure}
We show that the model ${H_{\pi}}$ with ${H_\text{A}=H'_{\text{A}}(\pi)}$ given by Eq.~\eqref{eq:annihilatepi} with parameters
\begin{equation}\label{eq:multipimagnon}
    \begin{aligned}
    H_{\pi}:  (p_0,J_0,J_1,J_2,J_3,\Omega)&=(\pi,0,0,0.1,1,2)\\
        (J^{\text{DM2}}_{x},J^{\text{DM2}}_{y},J^{\text{DM2}}_{z})&=(1.1,1.2,1.1)\\
        (J^{\text{RT}}_1,J^{\text{RT}}_2,J^{\text{RT}}_3,J^{\text{RT}}_4,J^{\text{RT}}_5)&=(1.2,2,1.9,1,1,1)
    \end{aligned}
\end{equation}
is scarred by the entire set of ${\{\psi_{n,N}(\pi)\}}$ states. The coefficients are meant to be generic and gives a well thermalizing spectrum at small system size.\par
\paragraph{Level statistics.} Fig. \ref{fig:prpi} shows that the model is thermalizing outside of the scarred manifold using the level spacing ratio. With the inclusion of multiple RT terms, ${H_{\pi}}$ does not possess any anti-unitary symmetry, thus the level statistics follow the Gaussian Unitary Ensemble (GUE) rather than GOE.\par

\paragraph{Entanglement entropy.} Fig. \ref{fig:piSE} shows that there are outlying eigenstates with subthermal entanglement entropy. They correspond to the ${p=0}$ and ${p_0=\pi}$ magnon states. Since states with the same total magnetization but different numbers of ${p=0,\pi}$ magnons are degenerate i.e.
\begin{equation}
\begin{aligned}
    H_{\pi}\ket{\psi_{n,N}(\pi)}&=E_{n,N}\ket{\psi_{n,N}(\pi)},\\
    E_{n,N}&=E_{0,0}-\Omega(n+N),
\end{aligned}    
\end{equation}
we confirm the identity of these outliers as follows. In the plot, the marked states ${\{E_n\}}$ constitute a scarred manifold of dimension ${d_{\text{scar}}=(L/2+1)^2}$. On the other hand, the set of magnon states in Eq.~\eqref{eq:multimagnon} can be orthonormalized to form a basis ${\{\ket{\tilde{\psi}_{n'}}\}}$ of dimension ${d_{\text{magnon}}}$. We verify that this basis spans the scarred manifold by checking that the projectors \begin{equation}
    P_\text{scar}=\sum_{n=1}^{d_\text{scar}}\ket{E_n}\bra{E_n},\quad P_{\text{magnon}}=\sum_{n'=1}^{d_{\text{magnon}}}\ket{\tilde{\psi}_{n'}}\bra{\tilde{\psi}_{n'}}
\end{equation}
satisfy
\begin{equation}
    \Tr{P_{\text{scar}}P_\text{magnon}}=d_{\text{scar}}=d_{\text{magnon}},
\end{equation}
and that
\begin{equation}
    (I-P_{\text{scar}})P_{\text{magnon}}=0,
\end{equation}
which shows that the complement of the scarred manifold is disjoint from the set of magnon states.
\par

\paragraph{Quench dynamics and revival.}

As in Sec.~\ref{sec:2pi3magnon}, the ${p_0=\pi}$ magnon scarred states can be probed with simple initial states. Notice that the product state 
\begin{equation}
\begin{aligned}
    \ket{\mathbb{Z}_{2}}&=\ket{\uparrow\downarrow\uparrow\downarrow\cdots\uparrow\downarrow}\\
    &=\sum_{n+N=L/2}a_{n,N}\ket{\psi_{n,N}(\pi)}
\end{aligned}
\end{equation}
is a linear combination of degenerate scarred states and has spatial periodicity 2. To observe periodic revivals, we require initial states with overlap on non-degenerate scarred states. We consider
\begin{equation}
    \ket{\Psi_1}=\mathcal{N}_1\exp{cS^-}\ket{\mathbb{Z}_2},
\end{equation}
where ${\mathcal{N}_1}$ is the normalization constant. Under time-evolution by ${H_{\pi}}$, the state remains entirely within the scarred manifold. One can show that
\begin{equation}
\begin{aligned}
    \ket{\Psi_1(t)}&=\mathcal{N}_1e^{-iE_Zt}\exp{ce^{i\Omega t}S^-}\ket{\mathbb{Z}_2}\\   
    &=\mathcal{N}_1e^{-iE_Zt}\prod_{j=0}^{L/2-1}\bigg(\ket{\uparrow}_{2j}+ce^{i\Omega t}\ket{\downarrow}_{2j}\bigg)\ket{\downarrow}_{2j+1}.\\ 
\end{aligned}
\end{equation}
where ${E_{Z}}$ is the energy of ${\ket{\mathbb{Z}_2}}$. Note that unlike ${\ket{\Phi(t)}}$ in Eq.~\eqref{eq:phiT}, ${\ket{\Psi_1(t)}}$ remains a product state in real space. 

Hence, defining the fidelity $\mathcal{F}_1(t)$ of $|\Psi_1(t)\rangle$  as 
\begin{equation} \label{eq:F1def}
    \begin{aligned}
        \mathcal{F}_1(t)&=|\braket{\Psi_1}{\Psi_1(t)}|^2,
    \end{aligned}
\end{equation}
we have that under time evolution by $H_\pi$ that 
\begin{equation}
    \begin{aligned}
        \mathcal{F}_1(t)=\bigg(\frac{1+|c|^4+2|c|^2\cos\Omega t}{1+|c|^4+2|c|^2}\bigg)^{L/2}.
    \end{aligned}
\end{equation}

\subsection{Deforming the multi-${\pi}$-magnon model}\label{sec:deformpipi}

The nearest-neighbor Heisenberg interaction ${H^{\text{H}}_1}$ is an experimentally relevant deformation. We consider
\begin{equation}
H_{\pi}(\epsilon)=H_{\pi}+\epsilon H^{\text{H}}_{1}
\end{equation}
and show that the fidelity of the initial state ${\ket{\Psi_1(t)}}$ no longer exhibits perfect revivals. Furthermore, there are no revivals in the fidelity at late times in the limit ${L\to\infty}$. 
\par
\begin{figure}
\includegraphics[width=\linewidth]{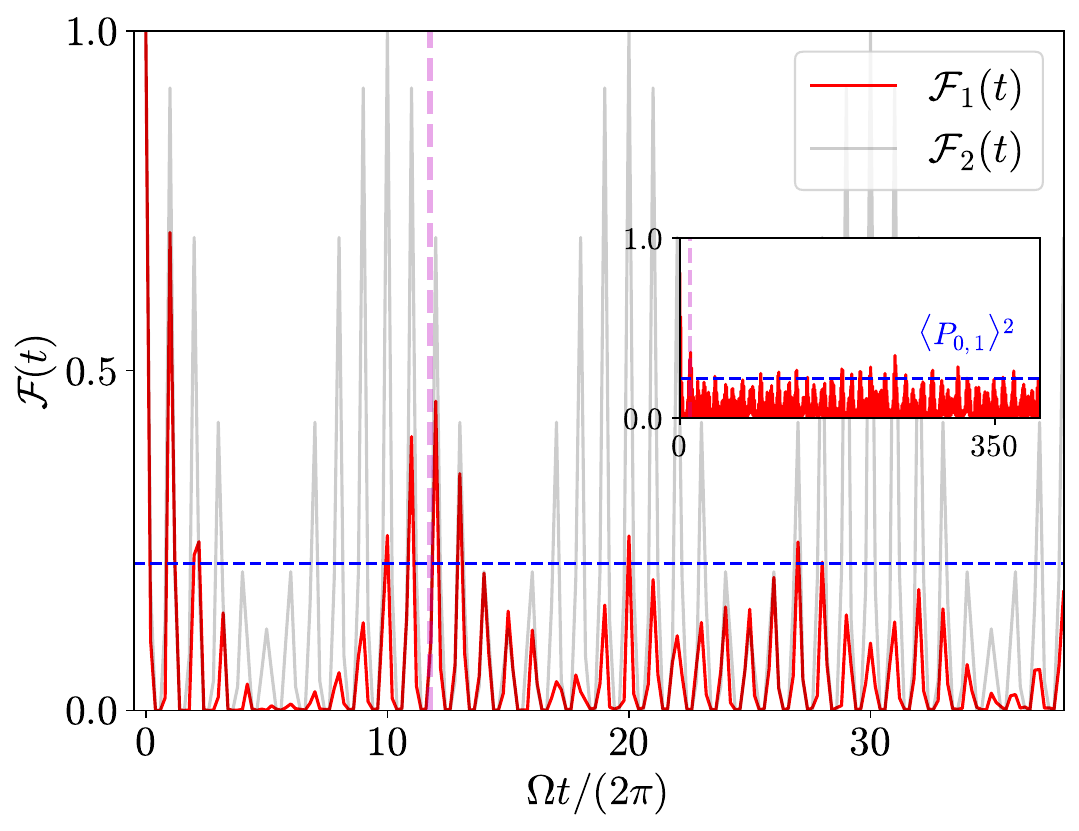}

		\caption{${\mathcal{F}_1(t)}$ (defined in Eq. \eqref{eq:F1def} and  plotted in red) shows persistent oscillations which decay to an amplitude set by ${\sim\langle P_{0,1}\rangle^2}$ (the horizontal blue dashed line, defined in Eq. \eqref{eq:projexpec} and computed in Eq.~\eqref{eq:largeLP}) after time ${t\sim2\pi/(\epsilon (\delta H^{\text{H}}_1))}$ (vertical pink dashed line, with ${\delta H_1^{\text{H}}}$ defined in Eq.~\eqref{eq:energywidth}). In contrast, the quantity ${\mathcal{F}_2(t)/\langle P_{0,1}\rangle^2}$ (defined in Eq. \eqref{eq:projfidel} and plotted in gray), normalized by $\langle P_{0,1}\rangle^2$ to avoid obstruction by the red line, shows perfect revivals with period  ${\Omega/E_{\pi}=10}$. The inset is a zoomed out version of the main plot. (${\epsilon=0.1,c=1, L=12}$)}

\label{fig:main3}

\end{figure}
The nearest-neighbor Heisenberg interaction ${H^{\text{H}}_1}$ only preserves the scars ${\{\ket{\psi_{n,N}(\pi)}\}_{N=0,1}}$ because
\begin{equation}\label{eq:Heisenstate}
\begin{aligned}
    &\epsilon H^{\text{H}}_{1}\big(Q^{-}(\pi)\big)^N\ket{\text{up}}\\
    &=(E_0-NE_{\pi})\big(Q^{-}(\pi)\big)^N\ket{\text{up}}\\
    &-2N(N-1)\epsilon\big(Q^{-}(\pi)\big)^{N-2}\sum_{i=0}^{L-1}S^-_{i}S^-_{i+1}\ket{\text{up}}
\end{aligned}
\end{equation}
where ${H_{1}^{H}\ket{\text{up}}=E_0\ket{\text{up}}=(\epsilon L/4)\ket{\text{up}}}$ and ${E_\pi=2\epsilon}$ is the energy of a single ${p_0=\pi}$ magnon. The interaction connects ${N>1}$ ${p_0=\pi}$ magnon states to other generic states in the spectrum of ${H_{\pi}(\epsilon)}$.\par 

At short times, the fidelity ${\mathcal{F}_1(t)=|\braket{\Psi_1}{\Psi_1(t)}|^2}$ oscillates as it would with ${\epsilon=0}$. It then decays due to the overlap with non-scarred eigenstates. A conservative estimate for the timescale at which we expect to see decay is set by ${T_d=2\pi/(\epsilon\delta H^{\text{H}}_1)}$, where the energy width
\begin{equation}\label{eq:energywidth}
    \delta H^{\text{H}}_{1}=\text{max}\bigg\{\sqrt{\langle (H^{\text{H}}_{1})^2\rangle_{n,N}-\langle H^{\text{H}}_{1}\rangle_{n,N}^2}\bigg\}=\order{L^{1/2}}
\end{equation}
is maximized with respect to ${n,N}$ of the multi-magnon states ${\ket{\psi_{n,N}(\pi)}}$ for a given ${L}$. In Fig.~\ref{fig:main3}, ${t=T_d}$ is plotted as a vertical pink dashed line.\par 

At ${t>T_d}$, fidelity oscillations persist with reduced amplitude. This is a consequence of the finite overlap of ${\ket{\Psi_1(t)}}$ with the ${N=0,1}$ ${p=\pi}$ magnon scarred manifold at finite size. Consider the projector onto the exact scarred manifold
\begin{equation}
\begin{aligned}
P_{0,1}=\sum_{N=0,1}\sum_{n}\frac{\ket{\psi_{n,N}(\pi)}\bra{\psi_{n,N}(\pi)}}{\braket{\psi_{n,N}(\pi)}{\psi_{n,N}(\pi)}},
\end{aligned}
\end{equation}
as well as the fidelities of the projections
 \begin{equation}\label{eq:projfidel}
     \mathcal{F}_2(t)=|\bra{\Psi_1(t)}P_{0,1}\ket{\Psi_1}|^2,
 \end{equation}
 and 
  \begin{equation}\label{eq:projfidel3}
     \mathcal{F}_3(t)=|\bra{\Psi_1(t)}(I-P_{0,1})\ket{\Psi_1}|^2.
 \end{equation}
The maximal value of ${ \mathcal{F}_2(t)}$ is given by ${\langle P_{0,1}\rangle^2}$, where the expectation value is given by
 \begin{equation}\label{eq:projexpec}
     \langle P_{0,1}\rangle=\bra{\Psi_1}P_{0,1}\ket{\Psi_1}=\bra{\Psi_1(t)}P_{0,1}\ket{\Psi_1(t)}.
 \end{equation}
%
In Fig.~\ref{fig:main3}, ${\langle P_{0,1}\rangle^2}$ is plotted as a horizontal blue dashed line. At late times, the amplitude of ${\mathcal{F}_1(t)}$ decays to a value approximately given by ${\langle P_{0,1}\rangle^2}$.\par 

\begin{figure}
\includegraphics[width=\linewidth]{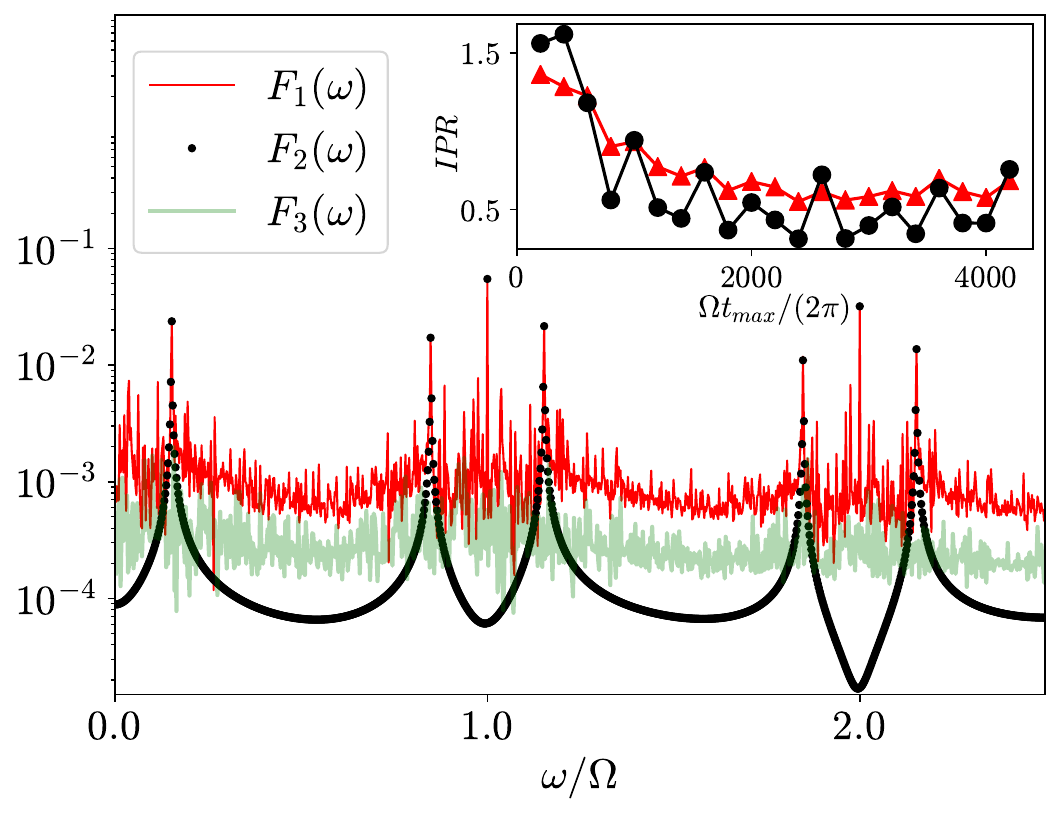}

		\caption{ ${F_1(\omega),F_2(\omega)}$ and ${F_3(\omega)}$ are respectively the Fourier spectra of ${\mathcal{F}_1(t)}$, ${\mathcal{F}_2(t)}$ and ${\mathcal{F}_3(t)}$ (Eqs. \eqref{eq:F1def}, \eqref{eq:projfidel}, \eqref{eq:projfidel3}) at late times. ${F_3(\omega)}$ is featureless. On the other hand, both ${F_1(\omega)}$ and ${F_2(\omega)}$ show sharp peaks at ${\omega=|n\Omega\pm E_{\pi}|}$ and at ${\omega=n\Omega}$ with equal amplitude, indicating that the coherent oscillations of ${\mathcal{F}_1(t)}$ are due to overlap with the ${N=0,1}$ ${p=\pi}$ magnon scarred states. The inset shows the IPR (Eq.~\eqref{eq:IPR}) with respect to ${F_1(\omega)}$ (triangles) and ${F_2(\omega)}$ (circles). Its saturation at increasing total simulation time ${t_{\text{max}}}$ is an indication of the localization of frequency distribution (at ${\omega=n\Omega}$ and ${\omega=|n\Omega\pm E_{\pi}|}$). The Fourier spectra are plotted at such late times. (${\epsilon=1.1\pi/3,c=1 ,L=12}$) }
\label{fig:main4}
\end{figure}

 Fig.~\ref{fig:main4} shows the Fourier spectra of ${\mathcal{F}_1(t)}$, ${\mathcal{F}_2(t)}$, and ${\mathcal{F}_3(t)}$ at late times (denoted by ${F_1(\omega)}$, ${F_2(\omega)}$, and ${F_3(\omega)}$ respectively). For a given ${n<L-1}$, ${\ket{\psi_{n,1}(\pi)}}$ evolves with an extra phase factor from ${E_{\pi}}$ as compared to ${\ket{\psi_{n,0}(\pi)}}$. Thus, ${F_2(\omega)}$ has sharp peaks at ${\omega=n\Omega}$ and ${\omega=|n\Omega\pm E_{\pi}|}$. The peaks of ${F_1(\omega)}$ match those of ${F_2(\omega)}$, while the Fourier spectrum of ${\mathcal{F}_3(t)}$ is featureless. We further probe the localization of frequency distribution around these peaks with the inverse participation ratio (IPR). For a given set of ${\{F(\omega_j)\}}$ generated from time evolution up to total simulation time ${t_{\text{max}}}$, the IPR may be computed as 
 \begin{equation}\label{eq:IPR}
     \text{IPR}=\frac{R_4}{(R_2)^2},\quad R_n=\frac{\sum_{j} \omega_j^n F(\omega_j)}{\sum_{j} F(\omega_j)}. 
 \end{equation}
 These plots indicate that the late-time coherent oscillation of ${\mathcal{F}_1(t)}$ entirely stems from the overlap with the ${N=0,1}$ ${p=\pi}$ magnon scarred states.\par 
 
 As a side note, the sharp peaks at ${\omega=|n\Omega\pm E_{\pi}|}$ are not Delta-like (which is the case at ${\omega=n\Omega}$) because for incommensurate ${E_{\pi}}$ and ${\Omega}$, the Fourier expansion of ${\mathcal{F}_2(t)}$ has weight on all frequencies. In addition, a larger ${\epsilon}$ (${\approx 1.15}$) is used in Fig.~\ref{fig:main4} to produce a more uniform ${F_3(\omega)}$. At ${\epsilon=0.1}$ (as in Fig.~\ref{fig:main3}), ${F_3(\omega)}$ still has small and relatively smooth local maxima around ${\omega=n\Omega}$, which becomes increasingly uniform at larger ${\epsilon}$.\par

In the limit ${L\gg1}$, the expectation value of the projector, which sets the amplitude of the oscillations in ${\mathcal{F}_1(t)}$ at late times, decreases exponentially with system size ${L}$ as 

\begin{equation}\label{eq:largeLP}
\begin{aligned}
\langle P_{0,1}\rangle=&\sum_{N=0,1}\sum_{n+N\leq L/2}\bigg(\frac{|c|^{L/2-N-n}(L/2)!}{(L/2-N-n)!}\bigg)^2\\
&\times\frac{(L-2N-n)!}{L^Nn!(L-2N)!}\frac{1}{(1+|c|^2)^{L/2}}\\
\sim& 2^{-0.22L}\quad(c=1).
\end{aligned}
\end{equation}
Therefore, we do not expect to see any residual coherent oscillation even when the deformation preserves scars with zero or one ${p_0=\pi}$ magnon.\par 

The analysis in the case that the model is also deformed by nearest-neighbor Dzyaloshinskii-Moriya interactions (DM1) is analogous. After a timescale set by the energy width of the magnon states, the amplitude of ${\mathcal{F}_1(t)}$ decays to a value set of the expectation value of the projector onto the ${p=0}$ magnon scarred manifold, which decreases exponentially with ${L}$.

\subsection{Multi-${p_0}$-magnon model}\label{sec:multipnumerics}
 Finally, we show that the model described by Eq.~\eqref{eq:multipHam} with parameters 
\begin{equation}\label{eq:MMparameters}
    H^{\text{SM}}_{2\pi/3}: (p_0,M,J_M,K_1,\Omega)=(2\pi/3,3,1,1,1)
\end{equation}
is scarred by the entire set of states ${\{\psi_{n,N}(p_0=2\pi/3)\}}$. The scalar term ${H_{\text{sym}}}$ is added to shift the energy of the scarred states relative to the rest of the spectrum.

The model has two symmetries: translational invariance and U(1) rotations about the $z$-axis. We focus on the zero momentum and $S^z=2$ z-magnetization sector.

\begin{figure}
\includegraphics[width=\linewidth]{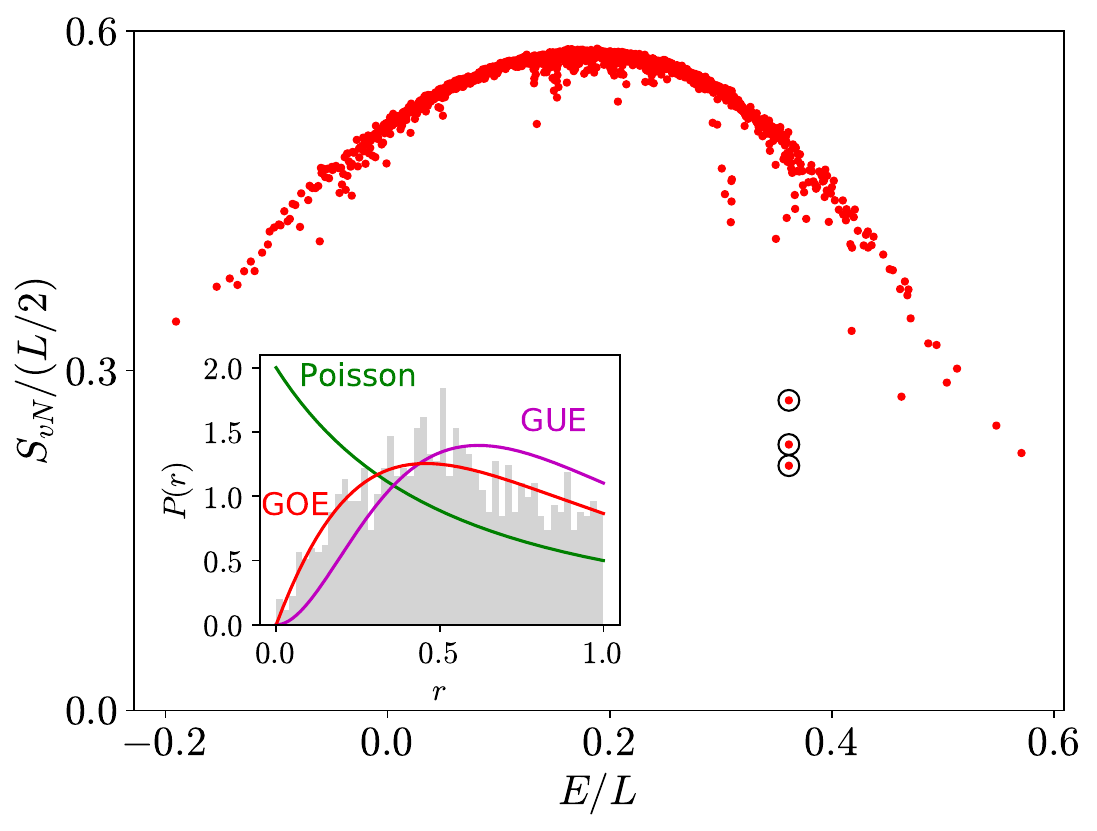}
	\caption{The main plot shows the half-chain entanglement entropy density plotted against the energy density of each eigenstate of ${H^{\text{SM}}_{2\pi/3}}$ (parameters given in Eq. \eqref{eq:MMparameters}, with $L=18$) in the ${p=0}$ momentum and ${S^z=2}$ magnetization sector. The three degenerate scarred states are marked by circles. The inset shows a normalized histogram of the level spacings ${r_n}$, closely matching the GOE prediction in \cite{atas_distribution_2013}. The mean value ${\langle r\rangle}$ in our data is ${\approx 0.531}$. }
\label{fig:main5}

\end{figure}

\paragraph{Level statistics.} The inset of Fig.~\ref{fig:main5} shows that the nearest neighbor energy level statistics follows that of the GOE at system size ${L=18}$.\par

\paragraph{Entanglement entropy.} The main plot in Fig.~\ref{fig:main5} shows three outlying eigenstates with distinctly subthermal half chain entanglement entropy. Since the multi-magnon state ${\ket{\psi_{n,N}(p_0)}}$ has total momentum ${Np_0}$ (mod ${2\pi}$) and magnetization ${L/2-(N+n)}$, the outliers can be identifed as the ${(N,n)=(0,7)}$, ${(N,n)=(3,4)}$ and ${(N,n)=(6,1)}$ orthogonal states. \par

\section{Discussion}\label{sec:conclusion}

In summary, irreducible tensor operators provide a natural basis to derive Hamiltonians with towers (or a pyramid) of quantum many-body scars. These Hamiltonians generally fall outside the Shiraishi-Mori formalism, and satisfy a restricted spectrum generating algebra. Indeed, several previous models of scars~\cite{mark__2020, schecter_weak_2019} can be interpreted as a linear combination of $Q$-SU(2) tensor operators (Appendix~\ref{sec:multipole}). (Three-spin) tensor operators may be realized in various quantum simulators~\cite{andrade_engineering_2022}. For instance, three-spin scalar operator of the form Eq.~\eqref{eq:scalarproduct} was synthesized in a system of three superconducting qubits using Floquet engineering~\cite{liu_synthesizing_2020}.\par

As an application, we have derived families of Hamiltonians scarred by multi-magnon states, which are built from the repeated actions of ladder operators ${Q^-(p_0)}$ and ${S^-}$ on the fully polarized state for any $p_0$.  Such states are generally neither eigenstates of the associated Casimir operators ${Q^2(p_0)}$ nor ${S^2}$. Generalized AKLT models provide the only other examples which realize the same feature in a single scarred tower. A natural extension of our work would be to construct AKLT-type Hamiltonians with multiple species of bi-magnon or magnon quasiparticle excitations atop the AKLT ground state, and Hamiltonians scarred by magnons with three or more quasi-momenta. \par 

We have primarily considered tensor operators of spin-SU(2). However, as discussed in Appendix~\ref{sec:multipole}, tensor operators can be constructed from other symmetries. Such symmetries need not arise from spin degrees of freedom nor be defined on single sites. This opens the possibility of embedding (multi-species) scars with multi-site quasiparticles, and/or scars in electronic models such as those with ${\eta}$-pairing SU(2) symmetry~\cite{mark__2020,moudgalya_rsga_2020}. \par 

In the multi-${\pi}$-magnon model ${H_{\pi}}$, many terms are tensor operators of both spin-SU(2) and ${Q(\pi)}$-SU(2) symmetry with the same ranks and components. The rest are either ${Q}$-SU(2) tensors with a different rank, or not well-defined tensors at all and only annihilate the base state by accident. Given this, a general principle for embedding scars with multiple species of quasiparticles (e.g. magnons with more than two different momenta) may be to use tensor operators associated with multiple symmetries.\par 

Furthermore, in several example Hamiltonians in the manuscript, individual spherical tensors annihilate the scarred tower. As tensor operators (with ${q\neq0}$) are non-Hermitian, the tensor operator formalism may be useful for embedding scars in open systems~\cite{buca_non-stationary_2019}.\par


\section{Acknowledgements}
The authors are grateful to Vedika Khemani, Sanjay Moudgalya, and Andrei Bernevig for helpful discussions. This work was supported by the National Science Foundation through the award DMR-1752759 (L.-H.T and A.C.), and by the US Department of Energy, Office of Science, Basic Energy Sciences, under the Early Career Award No.~DE-SC0021111 (N.O.D). This work was also performed in part at the Aspen Center for Physics, which is supported by National Science Foundation grant PHY-1607611. Numerical calculations were performed with the help of QuSpin \cite{weinberg_quspin_2017,weinberg_quspin_2019}.

\newpage
\appendix

\section{Derivation of the spherical tensor operator basis}\label{sec:completebasis}

Here, we explicitly construct the exhaustive list of spherical tensor operators composed of up to three spin-1/2 operators. These operators form a complete basis from which we derive scarred models, as discussed in Sec.~\ref{sec:tensorrsga}.\par 

The procedure is straightforward. Since the transformation properties of tensor operators under commutations with ladder operators is identical to that of applying ladder operators to angular momentum eigenstates (see Eqs.~\eqref{eq:STalgebra} and \eqref{eq:rank}), we may identify the trace-orthonormal one-body spherical tensor operators
\begin{equation}\label{eq:tensorstate}
    \begin{aligned}
    T^{q_i=0}_{(k_i=1)}=\sqrt{2}S^{z}_{i}&\leftrightarrow\ket{k_i=1,q_i=0}\\
    T^{q_i=\pm1}_{(k_i=1)}=\mp S^{\pm}_{i}&\leftrightarrow\ket{k_i=1,q_i=\pm1}.
    \end{aligned}
\end{equation}
Then, multi-site spherical tensor operators can be systematically combined using Clebsch Gordan (CG) coefficients, much like combining two angular momentum states. That is, a spherical tensor of rank ${\mathcal{K}_1}$ and component ${\mathcal{Q}_1}$ is obtained from 
\begin{equation}\label{eq:composite}
\begin{aligned}
    T^{\mathcal{Q}_1}_{(\mathcal{K}_1)}=&\sum_{q_1,q_2}\braket{k_1,q_1;k_2,q_2}{\mathcal{K}_1,\mathcal{Q}_1}T^{q_1}_{(k_1)}\otimes T^{q_2}_{(k_2)}
\end{aligned}
\end{equation}
where ${\braket{k_1,q_1;k_2,q_2}{\mathcal{K}_1,\mathcal{Q}_1}}$ is a CG coefficient in the expansion of the angular momentum state
\begin{equation}\label{eq:compositestate}
    \ket{\mathcal{K}_1,\mathcal{Q}_1}=\sum_{q_1,q_2}\braket{k_1,q_1;k_2,q_2}{\mathcal{K}_1,\mathcal{Q}_1}\ket{k_1,q_1}\otimes\ket{k_2,q_2}.
\end{equation}
These relations between states and tensor operators hold in greater generality, as discussed in Appendix~\ref{sec:multipole}. Following this, we briefly discuss why only certain representations of the tensor operators appear in the basis.\par

In the rest of this section, product states such as ${\ket{k_i=1,q_i}\otimes \ket{k_j=1,q_j}}$ are denoted by ${\ket{q_iq_j}}$ for brevity, where ${q_{i,j}=0,\pm}$.\par 

\subsection{Two-body operators}

The tensor product of spin operators on two different sites form the irreducible representations ${3\otimes3=1\oplus3\oplus5}$. It means the decomposition yields three sets of tensor operators with ranks ${K=0,1}$ or ${2}$; the numbers represent the corresponding dimensions ${2K+1}$.\par

${K=0}$: From
\begin{equation}\label{eq:cg110}
    \ket{K=0,Q=0}=\frac{1}{\sqrt{3}}(\ket{-+}+\ket{+-}-\ket{00}),
\end{equation}
we can identify states with operators via Eq.~\eqref{eq:tensorstate} to obtain the Heisenberg interaction in Eq.~\eqref{eq:heisenberg} as the two-body rank-0 (scalar) operator
\begin{equation}
    T^{Q=0}_{(K=0)}=\frac{-2}{\sqrt{3}}\boldsymbol{S}_{i}\cdot\boldsymbol{S}_{j}.
\end{equation}

${K=1}$: From
\begin{equation}
    \ket{K=1,Q=0}=\frac{1}{\sqrt{2}}(\ket{+-}-\ket{-+}),
\end{equation}
we obtain the Dzyaloshinskii-Moriya interaction in Eq.~\eqref{eq:DMR} as the ${Q=0}$ component of the two-body rank-1 (vector) operator
\begin{equation}
        T^{Q=0}_{(K=1)}=\frac{i}{\sqrt{2}}(\hat{z}\cdot\boldsymbol{S}_i\times\boldsymbol{S}_{j}).
\end{equation}
The ${Q=\pm 1}$ components can either by obtained from angular momentum eigenstates or by commuting ${T^{Q=0}_{(K=1)}}$ with the total raising/lowering operator ${S^{\pm}_{i}+S^{\pm}_{j}}$. Note that the overall (imaginary) coefficient can be set to unity in the operator basis.\par 

${K=2}$: From
\begin{equation}
    \ket{K=2,Q=0}=\frac{1}{\sqrt{6}}(\ket{+-}+\ket{-+}+2\ket{00}),
\end{equation}
we obtain the ${Q=0}$ component of the two-body rank-2 operator
\begin{equation}\label{eq:twobodyrank2a}
    T^{Q=0}_{(K=2)}=\sqrt{\frac{2}{3}}(3S^z_{i}S^z_{j}-\boldsymbol{S}_{i}\cdot\boldsymbol{S}_{j}).
\end{equation}
The rest of the components can be obtained similarly. Since ${K=k_i+k_j=2}$ is the highest possible rank, it is just as easy to start from the lowest-weight operator ${T^{Q=-2}_{(K=2)}\propto S^{-}_{i}S^-_{j}}$.

\subsection{Three-body operators}

Three-body terms are obtained by forming irreducible representations from the two-site representations discussed above and a ${k_3=1}$ representation on a third site. They can be decomposed into ${3\otimes(1\oplus3\oplus5)=3\oplus (1\oplus3\oplus5)\oplus(3\oplus5\oplus7)}$, i.e. one scalar, three sets of vectors, two sets of rank-2 tensors and one set of rank-3 tensors. \par

${K'=0}$: Using the same CG coefficients in Eq.~\eqref{eq:cg110}, we combine a one-body triplet (${k_l=1}$) and the two-body triplet (${K=1}$) constructed above as 
\begin{equation}
\frac{\ket{k_l,-1}\otimes\ket{K,1}+\ket{k_l,1}\otimes\ket{K,-1}-\ket{k_l,0}\otimes\ket{K,0}}{\sqrt{3}}
\end{equation}
which yields the scalar triple product term in Eq.~\eqref{eq:scalarproduct}
\begin{equation}
    T^{Q=0}_{(K'=0)}\propto \boldsymbol{S}_{l}\cdot(\boldsymbol{S}_{i}\times\boldsymbol{S}_{k}).
\end{equation}

${K'=1}$: There are three sets of vectors, one from each of the ${K=0,1,2}$ two-body tensor operators. Using the CG coeficients for ${3\otimes1\to 3}$, ${3\otimes3\to3}$ and ${3\otimes 5\to 3}$, we obtain
\begin{equation}\label{eq:3br1a}
    T^{Q=0}_{(K'=1),\phi_1=0}\propto S^z_{l}(\boldsymbol{S}_{i}\cdot\boldsymbol{S}_{j})
\end{equation}
\begin{equation}\label{eq:3br1b}
    T^{Q=0}_{(K'=1),\phi_1=1}\propto S^z_{i}(\boldsymbol{S}_{l}\cdot\boldsymbol{S}_{j})-S^z_{j}(\boldsymbol{S}_{l}\cdot\boldsymbol{S}_{i})
\end{equation}
\begin{equation}\label{eq:3br1c}
\begin{aligned}
    T^{Q=0}_{(K'=1),\phi_1=2}&\propto \sqrt{\frac{3}{4}}\bigg(S^z_{i}(\boldsymbol{S}_{l}\cdot\boldsymbol{S}_{j})+S^z_{j}(\boldsymbol{S}_{l}\cdot\boldsymbol{S}_{i})\bigg)\\
    &-\frac{S^z_{l}(\boldsymbol{S}_{i}\cdot\boldsymbol{S}_{j})}{\sqrt{3}},
\end{aligned}
\end{equation}
where ${\phi_1}$ labels the different representations of rank-1 tensors. The values of ${\phi_1}$ are chosen to be the same as ${K}$, the rank of the two-body tensor operator from which the representation is formed. It is straightforward to see that the rest of the components are obtained simply by replacing the single-site ${S_r^z}$ operators with ${\mp S_r^{\pm}}$.\par 

${K'=2}$: Similarly, there are two sets of rank-2 operators from the  ${K=1,2}$ two-body tensor operators, given by
\begin{equation}\label{eq:rank2b}
\begin{aligned}
    T^{Q=\pm2}_{(K'=2),\phi_2=1}&= \pm(S^{\pm}_{l}S^{\pm}_{i}S^z_{j}-S^{\pm}_{l}S^z_{i}S^{\pm}_{j}).
\end{aligned}
\end{equation}
\begin{equation}\label{eq:rank2a}
\begin{aligned}
    T^{Q=\pm2}_{(K'=2),\phi_2=2}&=\frac{ 2S^z_{l}S^{\pm}_{i}S^{\pm}_{j}-S^{\pm}_{l}S^{\pm}_{i}S^z_{j}-S^{\pm}_{l}S^z_{i}S^{\pm}_{j}}{\sqrt{3}}
\end{aligned}
\end{equation}

${K'=3}$: Finally, as in the two-body case, since ${K=k_l+k_i+k_j=3}$ is the highest possible rank for three-body operators, there is only one set of rank-3 tensors with the extremal weight components given by
\begin{equation}
    T^{Q=\pm3}_{(K'=3)}=\mp S^{\pm}_{l}S^{\pm}_{i}S^{\pm}_{j}.
\end{equation}

\subsection{Forming the basis}

As discussed in Sec.~\ref{sec:utility}, given a highest-weight base state with fixed ${z}$-magnetization ${\ket{\psi_0}=\ket{\text{up}}=\bigotimes_{i}\ket{\uparrow}_{i}}$, we may derive tensor operators that appear in ${H_{\text{A}}}$ for each ${k}$ and ${q}$ separately. Since any operator that annihilates a multi-magnon state must also annihilate ${\ket{\text{up}}}$, the rationale is to use linear combinations of the operators above to construct a maximal set of tensor operators that individually annihilate ${\ket{\text{up}}}$ (starting from those with the lowest weight), and exclude the remaining ones.\par 

Take ${l=i-1,j=i+1}$ so that ${l<i<j}$ label three contiguous sites (this can be immediately generalized to ${l=i-R, j=i+R}$). We consider their translationally-invariant sums as defined in Eq.~\eqref{eq:sumtensor}.\par 

First, there are no rank-3 tensor operators in the basis because it is impossible to satisfy Eq.~\eqref{eq:explicitconstraint} with ${q-n=-3}$ with just one set of rank-3 tensor operators in the basis.
More generally, for our choice of ${\ket{\psi_0}}$, we may determine the whole set of ${c^{\phi_p}_{q}}$ for certain representations ${\phi_p}$ of a given rank ${p}$. Suppose ${T^{-p}_{(p),\phi_p}\ket{\psi_0}\neq0}$ only for a particular representation ${\phi_p}$. Then the coefficient for the entire set of tensors ${\{T^{q}_{(p),\phi_p}\}}$ must vanish i.e. ${c^{\phi_p}_q=0}$ for all ${q=-p,\cdots p}$. This can be shown by iteratively considering the ${n=2p,2p-1,\cdots}$ equations in Eq.~\eqref{eq:explicitconstraintwithmag}. For example, when ${n=2p}$, the ${q=p}$ component gives
\begin{equation}\label{eq:lowweight}
\begin{aligned}
    c^{\phi_{p}}_{p}T^{-p}_{(p),\phi_{p}}\ket{\psi_0}&=0\\
    T^{-p}_{(p),\phi_p}\ket{\psi_0}&\neq0\implies c_{p}^{\phi_p}=0=c_{-p}^{\phi_p}.
\end{aligned}
\end{equation}
Then, when ${n=2p-1}$, the ${q=p-1}$ component gives
\begin{equation}
\begin{aligned}
     c^{\phi_p}_{p-1}T^{-p}_{(p),\phi_p}\ket{\psi_0}&=0\\
    T^{-p}_{(p),\phi_p}\ket{\psi_0}&\neq0\implies c_{p-1}^{\phi_p}=0=c_{-p+1}^{\phi_p}, \label{Eq:HighestWeightZero}
\end{aligned}
\end{equation}
and so on, leading to ${c^{\phi_p}_{q}=0}$ for all ${q}$. In this specific instance, since the lowest-weight operator
\begin{equation}\label{eq:maxranktensor}
    T^{-K}_{(K)}=\sum_{i=0}^{L-1} S^-_{i+1}S^-_{i+2}\cdots S^-_{i+K} 
\end{equation}
does not annihilate ${\ket{\psi_0}=\ket{\text{up}}}$, ${H_{\text{A}}}$ has no support on the entire set of ${\{T^{q}_{(K)}\}}$ with ${K=3}$.\par


Second, note that the rank-2 tensor operators in Eq.~\eqref{eq:rank2a} and \eqref{eq:rank2b} can be combined to form

\begin{equation}
    T^{-2}_{(2),a}=\sum_{i=0}^{L-1}(S^z_{i-1}S^-_{i}S^-_{i+1}-S^-_{i-1}S^-_{i}S^z_{i+1})
\end{equation}

\begin{equation}
    T^{-2}_{(2),b}=\sum_{i=0}^{L-1}(S^z_{i-1}S^-_{i}S^-_{i+1}-2S^-_{i-1}S^z_{i}S^-_{i+1}+S^-_{i-1}S^-_{i}S^z_{i+1}).
\end{equation}

${T^{-2}_{(2),a}}$ annihilates ${\ket{\text{up}}}$ as a sum, as do the rest of ${T^{q}_{(2),a}}$ components. They therefore form part of the operator basis. On the other hand, neither ${T^{-2}_{(2),b}}$ nor ${T^{-1}_{(2),b}}$ annihilates ${\ket{\text{up}}}$. One then needs to determine if the other rank-2 tensors in the construction (i.e. those associated with Eq.~\eqref{eq:twobodyrank2a}) can be used to form linear combinations that do annihilate the state. This can be achieved by
\begin{equation}
    T^{-2}_{(2),c}=T^{-2}_{(2),b}-\sum_{i=0}^{L-1}(S^-_{i}S^-_{i+1}-S^-_{i-1}S^-_{i+1}).
\end{equation}
While the rest of the components also annihilate ${\ket{\text{up}}}$, we reject them since ${T^0_{(2),c}}$ is non-Hermitian. The other representations of rank-2 tensors are rejected for the same reason as the rank-3 tensors.\par

Third, the rank-1 tensors in Eqs.~\eqref{eq:3br1a}, \eqref{eq:3br1b} and \eqref{eq:3br1c} form two sets of operators whose components individually annihilate ${\ket{\text{up}}}$:

\begin{equation}
\begin{aligned}
    T^{-1}_{(1),a}=\sum_{i=0}^{L-1}&{S}_{i-1}^{-}(\boldsymbol{S}_{i}\cdot\boldsymbol{S}_{i+1})-2{S}^{-}_{i}(\boldsymbol{S}_{i-1}\cdot\boldsymbol{S}_{i+1})\\
    &+(\boldsymbol{S}_{i-1}\cdot\boldsymbol{S}_{i}){S}^{-}_{i+1},
\end{aligned}
\end{equation}
\begin{equation}
\begin{aligned}
    T^{-1}_{(1),b}=\sum_{i=0}^{L-1}&{S}_{i-1}^{-}(\boldsymbol{S}_{i}\cdot\boldsymbol{S}_{i+1})-(\boldsymbol{S}_{i-1}\cdot\boldsymbol{S}_{i}){S}^{-}_{i+1},
\end{aligned}
\end{equation}
and one set that does not, 
\begin{equation}
\begin{aligned}
    T^{-1}_{(1),c}=\sum_{i=0}^{L-1}&{S}_{i-1}^{-}(\boldsymbol{S}_{i}\cdot\boldsymbol{S}_{i+1})+{S}^{-}_{i}(\boldsymbol{S}_{i-1}\cdot\boldsymbol{S}_{i+1})\\
    &+(\boldsymbol{S}_{i-1}\cdot\boldsymbol{S}_{i}){S}^{-}_{i+1},
\end{aligned}
\end{equation}
which can be rejected.\par 

To summarize, we have effectively derived all the tensor operators in Sec.~\ref{sec:stbasis} which annihilate ${\ket{\text{up}}}$. These operators can be used to construct ${H_{\text{A}}}$ that annihilate multi-magnon states. This generally cumbersome process is simplified by two properties of the ${\ket{\text{up}}}$ state: 1. it is inversion symmetric, and 2. it is an eigenstate of $(\boldsymbol{S}_{i}\cdot\boldsymbol{S}_{j})$ for any $i,j$. 


\section{Actions of spherical tensors on base states}\label{sec:commutatorsbase}
For some of the spherical tensors $H$ defined in Sec. \ref{sec:stbasis}, we provide the analytical expressions of
\begin{equation}
    HQ^-(p)\ket{\boldsymbol{s}}=\big[H,Q^-(p)\big]\ket{\boldsymbol{s}}+Q^-(p)H\ket{\boldsymbol{s}},
\end{equation}
where ${\ket{\boldsymbol{s}}}$ and ${Q^-(p)}$ are the fully polarized state and magnon creation operator defined in Eq.~\eqref{eq:fullpol} and Eq.~\eqref{eq:magnonop} respectively.

Heisenberg interaction terms:
\begin{equation}
    \begin{aligned}
    &H^{\text{H}}_{R}\ket{\boldsymbol{s}}=Ls^2\ket{\boldsymbol{s}},\\
    &\big[H^{\text{H}}_{R},Q^-(p)\big]\ket{\boldsymbol{s}}=-4s\sin^2\frac{pR}{2}Q^-(p)\ket{\boldsymbol{s}}.
    \end{aligned}
\end{equation}
Scalar-triple-product terms:
\begin{equation}
    \begin{aligned}
    &H^{\text{ST}}\ket{\boldsymbol{s}}=0,\\
    &\big[H^{\text{ST}},Q^-(p)\big]\ket{\boldsymbol{s}}=-8s^2\sin(p)\sin^2\bigg(\frac{p}{2}\bigg)Q^-(p)\ket{\boldsymbol{s}}.
    \end{aligned}
\end{equation}
Dzyaloshinskii-Moriya interaction terms:
\begin{equation}
    \begin{aligned}
    &H_{R}^{\text{DM}}(\mu)\ket{\boldsymbol{s}}=0,\\
    &\big[H_{R}^{\text{DM}}(\mu),Q^-(p)\big]\ket{\boldsymbol{s}}\\
   =&-\frac{(\delta_{x\mu}+i\delta_{y\mu})}{2}e^{-i\frac{pR}{2}}2\sin\frac{pR}{2}\sum_{i=0}^{L-1}e^{-ipr_i}S^{-}_{i}S^{-}_{i+R}\ket{\boldsymbol{s}}\\
    &-\delta_{z\mu}2s\sin(pR)Q^{-}(p)\ket{\boldsymbol{s}}.
    \end{aligned}
\end{equation}
Vector-triple-product-plus terms:
\begin{equation}
    \begin{aligned}
    &H^{\text{VTP}}(\mu)\ket{\boldsymbol{s}}=0,\\
    &\big[H^{\text{VTP}}(\mu),Q^-(p)\big]\ket{\boldsymbol{s}}\\
=&\frac{(\delta_{x\mu}+i\delta_{y\mu})}{2}\sum_{i=0}^{L-1}e^{-ipr_i}\bigg[\bigg(4e^{-ip}s\sin^2\frac{p}{2}S^{-}_{i}S^{-}_{i+2}\\
&+4e^{-i\frac{p}{2}}s\sin\frac{p}{2}\sin({p})S^-_{i}S^-_{i+1}\bigg)\bigg]\ket{\boldsymbol{s}}\\
    &+\delta_{z\mu}4s^2[\cos(p)-\cos(2p)]Q^-(p)\ket{\boldsymbol{s}}.
    \end{aligned}
\end{equation}
Vector-triple-product-minus terms:

\begin{equation}
    \begin{aligned}
    &H^{\text{VTM}}(\mu)\ket{\boldsymbol{s}}=0,\\
    &\big[H^{\text{VTM}}(\mu),Q^-(p)\big]\ket{\boldsymbol{s}}\\
=&\frac{(\delta_{x\mu}+i\delta_{y\mu})}{2}\sum_{i=0}^{L-1}e^{-ipr_i}\bigg[\bigg(2ie^{-ip}s\sin(p) S^{-}_{i}S^{-}_{i+2}\\
&-4ie^{-i\frac{pR}{2}}s\sin\frac{p}{2}\cos({p})S^-_{i}S^-_{i+1}\bigg)\bigg]\\
    &-\delta_{z\mu}8s^2\sin^2\frac{p}{2}Q^-(p)\ket{\boldsymbol{s}}
    \end{aligned}
\end{equation}
The rank-2 terms formed from ${T^{q}_{(2),a}}$ in Eq.~\eqref{eq:RTTlist} can be expressed as

\begin{equation}
    \begin{aligned}
         H^{\text{RT}}(1)=\sum_{i=0}^{L-1}&\bigg[S^{x}_{i-1}(S^{y}_{i}S^{y}_{i+1}-S^{z}_{i}S^{z}_{i+1})\\
         &-(S^{y}_{i-1}S^{y}_{i}-S^{z}_{i-1}S^{z}_{i})S^{x}_{i+1}\bigg]
    \end{aligned}
\end{equation}
\begin{equation}
    \begin{aligned}
    H^{\text{RT}}(2)=\sum_{i=0}^{L-1}&\bigg[S^{y}_{i-1}(S^{z}_{i}S^{z}_{i+1}-S^{x}_{i}S^{x}_{i+1})\\
    &-(S^{z}_{i-1}S^{z}_{i}-S^{x}_{i-1}S^{x}_{i})S^{y}_{i+1}\bigg]
    \end{aligned}
\end{equation}
\begin{equation}
    \begin{aligned}
    H^{\text{RT}}(3)=\sum_{i=0}^{L-1}&\bigg[S^{z}_{i-1}(S^{x}_{i}S^{x}_{i+1}-S^{y}_{i}S^{y}_{i+1})\\
    &-(S^{x}_{i-1}S^{x}_{i}-S^{y}_{i-1}S^{y}_{i})S^{z}_{i+1}\bigg]
    \end{aligned}
\end{equation}
\begin{equation}
    \begin{aligned}
 H^{\text{RT}}(4)=\sum_{i=0}^{L-1}&\bigg[S^{x}_{i-1}(S^{y}_{i}S^{z}_{i+1}+S^{z}_{i}S^{y}_{i+1})\\
 &-(S^{y}_{i-1}S^{z}_{i}+S^{z}_{i-1}S^{y}_{i})S^{x}_{i+1}\\
&-S^{y}_{i-1}(S^{z}_{i}S^{x}_{i+1}+S^{x}_{i}S^{z}_{i+1})\\
&+(S^{z}_{i-1}S^{x}_{i}+S^{x}_{i-1}S^{z}_{i})S^{y}_{i+1}\bigg]\\
    \end{aligned}
\end{equation}
\begin{equation}
    \begin{aligned}
          H^{\text{RT}}(5)=\sum_{i=0}^{L-1}&\bigg[S^{z}_{i-1}(S^{x}_{i}S^{y}_{i+1}+S^{y}_{i}S^{x}_{i+1})\\
          &-(S^{x}_{i-1}S^{y}_{i}+S^{y}_{i-1}S^{x}_{i})S^{z}_{i+1}\bigg].
    \end{aligned}
\end{equation}
The actions of their commutators with ${Q^-(p)}$ on ${\ket{\boldsymbol{s}}}$ are omitted but they can be computed readily from ${T^{q}_{(2),a}}$.

\section{Derivation of ${V^{\mu}(p,s)}$}\label{sec:derivesol}
Here we present a derivation of the result in Eq.~\eqref{eq:vectorterm}. We assume ${H_{\text{A}}}$ only contains DM and VTP terms and solve for their coefficients using the equations Eq.~\eqref{eq:explicitconstraint} and the results in Appendix \ref{sec:commutatorsbase}.\par 

We proceed by letting
\begin{equation}
\begin{aligned}
    h_0&=H_{\text{A}}\\
    &=\sum_{\mu=x,y,z}\bigg[\alpha_{\mu}^{1}H^{\text{DM}}_{1}(\mu)+\alpha_{\mu}^{2}H^{\text{DM}}_{2}(\mu)+\alpha_{\mu}^{3}H^{\text{VTP}}(\mu)\bigg]\\
    &=\sum_{\phi_1=1,2,3}\bigg[\frac{\alpha^{\phi_1}_x+i\alpha^{\phi_1}_y}{\sqrt{2}}T_{(1),\phi_1}^{-1}\\
    &-\frac{(\alpha^{\phi_1}_x-i\alpha^{\phi_1}_y)}{\sqrt{2}}T_{(1),\phi_1}^{1}+\alpha^{\phi_1}_zT_{(1),\phi_1}^{0}\bigg].
\end{aligned}
\end{equation}
where ${\alpha^{\phi_k}_{\mu}}$ are real coefficients. Since ${H_{\text{A}}}$ in this case contains up to rank-1 tensors, we only need to consider
\begin{equation}
    \begin{aligned}
    &h_1=\big[H_{\text{A}},S^-\big]\\
    &=\sum_{\phi_1=1,2,3}\bigg[(\alpha^{\phi_1}_x-i\alpha^{\phi_1}_y)T_{(1),\phi_1}^{0}-\alpha^{\phi_1}_z\sqrt{2}T_{(1),\phi_1}^{-1}\bigg]
    \end{aligned}
\end{equation}
\begin{equation}
    \begin{aligned}
    h_2&=\bigg[\big[H_{\text{A}},S^-\big],S^-\bigg]\\
    &=\sum_{\phi_1=1,2,3}-\sqrt{2}(\alpha^{\phi_1}_x-i\alpha^{\phi_1}_y)T_{(1),\phi_1}^{-1}
    \end{aligned}
\end{equation}
as ${h_3=0}$.\par 

From Appendix \ref{sec:commutatorsbase}, we can read off

\begin{equation}
    T_{(1),\phi_1}^{1}Q^-(p)\ket{\boldsymbol{s}}=0
\end{equation}
\begin{equation}
    T_{(1),\phi_1}^{0}Q^-(p)\ket{\boldsymbol{s}}=\beta^{\phi_1}_0Q^-(p)\ket{\boldsymbol{s}}
\end{equation}
\begin{equation}
    T_{(1),\phi_1}^{-1}Q^-(p)\ket{\boldsymbol{s}}=(\beta^{\phi_1}_{-1,1}Q^-_1(p,p)+\beta^{\phi_1}_{-1,2}Q^-_2(p,p))\ket{\boldsymbol{s}}
\end{equation}
where the ${\beta^{\phi_1}}$ coefficients are all real and

\begin{equation}
    e^{-i\frac{p}{2}}\sum_{i=0}^{L=1}e^{-ipr_r}S^-_{i}S^{-}_{i+1}\ket{\boldsymbol{s}}\equiv Q^-_1(p,p)\ket{\boldsymbol{s}},
\end{equation}
\begin{equation}
    e^{-ip}\sum_{i=0}^{L=1}e^{-ipr_r}S^-_{i}S^{-}_{i+2}\ket{\boldsymbol{s}}\equiv Q^-_2(p,p)\ket{\boldsymbol{s}}
\end{equation}
are orthogonal to each other. From Eq.~\eqref{eq:explicitconstraint}, the ${n=0}$ equation yields
\begin{equation}\label{eq:decouple0}
    Q^-(p)\ket{\boldsymbol{s}}:\sum_{\phi_1=1}^{3}\alpha_{z}^{\phi_1}\beta_0^{\phi_1}=0
\end{equation}
\begin{equation}\label{eq:decouple1}
    \begin{aligned}        
        Q^-_{1}(p,p)\ket{\boldsymbol{s}}&:\sum_{\phi_1}(\alpha^{\phi_1}_x+i\alpha^{\phi_1}_y)\beta^{\phi_1}_{-1,1}=0\\    Q^-_{2}(p,p)\ket{\boldsymbol{s}}&:\sum_{\phi_1}(\alpha^{\phi_1}_x+i\alpha^{\phi_1}_y)\beta^{\phi_1}_{-1,2}=0.\\
    \end{aligned}
\end{equation}
Since the ${\alpha}$'s are real (by construction) and the ${\beta}$'s are also real (base-state dependent), we can separate the real and imaginary parts of Eq.~\eqref{eq:decouple1} into
\begin{equation}\label{eq:decouple2}
\begin{aligned}
    &\sum_{\phi_1}\alpha^{\phi_1}_x\beta^{\phi_1}_{-1,1}=0,\quad \sum_{\phi_1}\alpha^{\phi_1}_y\beta^{\phi_1}_{-1,1}=0,\\
    &\sum_{\phi_1}\alpha^{\phi_1}_x\beta^{\phi_1}_{-1,2}=0,\quad \sum_{\phi_1}\alpha^{\phi_1}_y\beta^{\phi_1}_{-1,2}=0.
\end{aligned}
\end{equation}
Likewise, the real and imaginary parts of the ${n=1}$ equation can be separated to yield
\begin{equation}\label{eq:decouple3}
\begin{aligned}
    &Q^-(p)\ket{\boldsymbol{s}}: \sum_{\phi_1}(\alpha^{\phi_1}_x-i\alpha^{\phi_1}_y)\beta_0^{\phi_1}\\
    &\implies \sum_{\phi_1}\alpha^{\phi_1}_x\beta_0^{\phi_1}=0,\quad \sum_{\phi_1}\alpha^{\phi_1}_y\beta_0^{\phi_1}=0
\end{aligned}
\end{equation}
\begin{equation}\label{eq:decouple4}
    \begin{aligned}        
        Q^-_{1}(p,p)\ket{\boldsymbol{s}}&:\sum_{\phi_1}\alpha^{\phi_1}_z\beta^{\phi_1}_{-1,1}=0,\\         Q^-_{2}(p,p)\ket{\boldsymbol{s}}&:\sum_{\phi_1}\alpha^{\phi_1}_z\beta^{\phi_1}_{-1,2}=0,\quad\\
    \end{aligned}
\end{equation}
Finally, there is the ${n=2}$ equation:
\begin{equation}\label{eq:decouple5}
    \begin{aligned}        
        Q^-_{1}(p,p)\ket{\boldsymbol{s}}&:\sum_{\phi_1}(\alpha^{\phi_1}_x-i\alpha^{\phi_1}_y)\beta^{\phi_1}_{-1,1}=0\\ Q^-_{2}(p,p)\ket{\boldsymbol{s}}&:\sum_{\phi_1}(\alpha^{\phi_1}_x-i\alpha^{\phi_1}_y)\beta^{\phi_1}_{-1,2}=0,\quad\\
    \end{aligned}
\end{equation}
Again, since the ${\alpha}$'s and ${\beta}$'s are real, this new equation is redundant, as the real and imaginary parts give the same equations as in Eq.~\eqref{eq:decouple2}.\par 
Having extracted the equations in terms of ${\beta}$'s, we can now solve for the \{${\alpha_{\mu}^{\phi_1}}$\}. For ${\mu=x,y}$, we have enough equations to solve for ${\alpha_{\mu}^{\phi_1}}$, since ${\phi_1=1,2,3}$, and Eq.~\eqref{eq:decouple2} and Eq.~\eqref{eq:decouple3} provide six equations. Noting that the equations for ${x}$ and ${y}$ are decoupled and are identical, we can write (for ${\mu=x,y}$)
\begin{equation}\label{eq:matsys}
   B\boldsymbol{\alpha}_{\mu}= \begin{bmatrix}
    \beta^{1}_{-1,1}&0&\beta^{3}_{-1,1}\\
    0&\beta^{2}_{-1,2}&\beta^{3}_{-1,2}\\
    \beta^{1}_{0}&\beta^{2}_{0}&\beta^{3}_{0}\\
    \end{bmatrix}\begin{bmatrix}
\alpha^{1}_{\mu}\\
\alpha^{2}_{\mu}\\
\alpha^{3}_{\mu}
    \end{bmatrix}
    =\boldsymbol{0}.
\end{equation}
From Appendix \ref{sec:commutatorsbase},
\begin{equation}
\begin{aligned}
    \beta^{1}_{-1,1}=-2\sin\frac{p}{2},&\quad \beta^{3}_{-1,1}=4s\sin\frac{p}{2}\sin{p},\\
        \beta^{2}_{-1,2}=-2\sin p,&\quad \beta^{3}_{-1,2}=4s\sin^2\frac{p}{2},\\
        \beta^{1}_{0}=-2s\sin p,\quad\beta^{2}_{0}&=-2s\sin 2p,\\
        \beta^{3}_{0}=4s^2(\cos{p}&-\cos{2p}),\\
\end{aligned}
\end{equation}
which can be used to compute
\begin{equation}
\begin{aligned}
    \det (B)&=\beta^{1}_{-1,1}\big(\beta^{2}_{-1,2}\beta^{3}_{0}-\beta^{2}_{0}\beta^{3}_{-1,2}\big)-\beta^{1}_{0}\beta^{2}_{-1,2}\beta^{3}_{-1,1}\\
    &=0.
\end{aligned}
\end{equation}
Thus the system permits non-trivial solutions. Using the first two rows, we obtain
\begin{equation}\label{eq:vecsol}
\begin{aligned}
    \alpha^1_{\mu}&=-\frac{\beta^3_{-1,1}}{\beta^{1}_{-1,1}}\alpha^3_{\mu}=(2s)\sin({p})\alpha^{3}_{\mu},\\ \alpha^{2}_{\mu}&=-\frac{\beta^3_{-1,2}}{\beta^{1}_{-1,2}}\alpha^3_{\mu}=s\tan\bigg(\frac{p}{2}\bigg)\alpha^{3}_{\mu}.
\end{aligned}
\end{equation}
Now, per Eq.~\eqref{eq:decouple4}, the first two rows in Eq.~\eqref{eq:matsys} are also satisfied by ${\mu=z}$, and the final row appears in part of Eq.~\eqref{eq:decouple0}. This implies Eq. \eqref{eq:vecsol} applies for ${\mu=x,y,z}$. Thus, the three vector components of the rank-1 tensors also decouple, resulting in Eq.~\eqref{eq:vectorterm}.\par 

In the special case ${p=\pi}$, ${\beta^{2}_{-1,2}}$, ${\beta^2_{0}}$, ${\beta^{3}_{-1,1}}$ and ${\beta^1_{0}}$ vanish. Hence Eq.~\eqref{eq:matsys} only permits a solution with ${\alpha^{1}_{\mu}=\alpha^{3}_{\mu}=0}$ and unconstrained ${\alpha^{2}_{\mu}}$ for ${\mu=x,y,z}$.\par 

\section{Spherical tensors beyond spin-1/2}\label{sec:spinsgen}

Since the procedure to obtain scarred Hamiltonians from spherical tensors follows from SU(2) symmetry, the results in Sec.~\ref{sec:magnonscar} are amenable to numerous generalizations.\par 

First, in a periodic spin-${s}$ chain, interpreting ${\{S^{\mu}_{i}\}}$ as the corresponding spin-${s}$ operator gives a family of spin-s models with multi-magnon states as scarred states (Sec.~\ref{sec:generalHam}). In addition, more tensor operators may be included in the basis for ${H}$. They are constructed from the generalized single-site spherical tensors  

\begin{equation}\label{eq:3jtensor}
\begin{aligned}
    &T^{q_i}_{(k_i)}(s)\\
    &=\sum_{m,m'}(-1)^{s-m}(2k_i+1)^{1/2}\begin{pmatrix}
  s & s & k_i \\
  m' & -m & q_i 
 \end{pmatrix}\ket{s,m}\bra{s,m'}\\
 &=\sum_{m,m'}(-1)^{s-m'}\braket{s,m;s,-m'}{k,q}\ket{s,m}\bra{s,m'}
\end{aligned}
\end{equation}
where ${m,m'=-s,-s+1,\cdots,s}$, ${k_i=0,1,\cdots 2s}$, ${q_i=-k_i,-k_i+1,\cdots k_i}$, and ${(\cdots)}$ is a Wigner 3-${j}$ symbol. Including the identity (${k_i=0}$), the set of ${(2s+1)^2}$ operators is complete in the single-site spin-${s}$ Hilbert space ${\mathcal{H}_{s}}$ and orthonormal with respect to the trace inner product. In particular, 
\begin{equation}
  T^{\pm k_i}_{(k_i)}\propto \big(S^{\pm}_{i}\big)^{k_i}.   
\end{equation}
The multi-site basis operators are then the irreducible representations obtained from Eq.~\eqref{eq:composite}.\par

Second, the family of models in Sec.~\ref{sec:generalHam} can also be immediately generalized by the ${Q}$-SU(2) operators 
\begin{equation}
\begin{aligned}
    &S^{\pm}_{i}\to Q^{\pm}_{i}=\frac{1}{(2s)!}\big(S^{\pm}_{i}\big)^{2s},\quad S^z_{i}\to Q^z_{i}=\frac{1}{2}\big[Q^+_{i},Q^-_{i}\big],\\
    &S^x_{i}\to Q^x_{i}=\frac{1}{2}\big(Q^+_{i}+Q^-_{i}\big),\quad S^y_{i}\to Q^y_{i}=\frac{1}{2i}\big(Q^+_{i}-Q^-_{i}\big),
\end{aligned}
\end{equation}
which only act nontrivially on the ${Q}$-spin-1/2 doublet ${\{\ket{s}_{i},\ket{-s}_i\}}$.  Therefore, with the replacement ${\{S^{+}_{i},S^-_{i},S^z_{i}\}\to\{Q^{+}_{i},Q^-_{i},Q^z_{i}\}}$, we find a family of operators that can be used to embed target states  
\begin{equation}
    \big(Q^-\big)^n\big(Q^-(p_0)\big)^N\ket{\boldsymbol{s}}
\end{equation}
built from magnon operators
\begin{equation}
    Q^{\pm}(p_0)=\frac{1}{(2s)!}\sum_{i=0}^{L-1}e^{\pm ipr_i}\big(S^{\pm}_{i}\big)^{2s},\quad Q^{\pm}(0)=Q^{\pm}.
\end{equation}
When ${s=1}$, ${Q^{\pm}(p_0)}$ are also known as bimagnon operators~\cite{schecter_weak_2019, mark__2020}. Note that as we discuss in Appendix~\ref{sec:multipole}, ${\{Q^+_{i},Q^-_{i},Q^z_{i}\}}$ do not form a complete basis of single-site ${Q}$-SU(2) spherical tensors for ${s>1/2}$.\par

\section{Tensor operators beyond SU(2)}\label{sec:multipole}

Below, we sketch a general method for generating complete bases of tensor operators of compact groups. \par 
\subsection{Construction of tensor operators}\label{sec:constr}
The properties of tensor operators are very similar to those of states transforming in an irreducible representation of the group. For example, such a state $|(k), q\rangle(S)$ can be labeled by its Hilbert space $S$ (specified by the number of sites and their local Hilbert space dimensions), a \emph{multi-indexed} $k$ which labels the representation, and a \emph{multi-indexed} $q$ which specifies the state within the representation. Under the action of a generator of the group acting on the space $S$, $Q_a(S)$, the state transforms as
\begin{equation}
    Q_a(S) |(k), q\rangle(S) = \sum_m (Q_{(k),a})_{mq} |(k), m\rangle(S)
\end{equation}
where $Q_{(k),a}$ is the $a$-th generator of the representation $k$.  Similarly, a tensor operator that acts on the space $S$ in the representation ${(k)}$ is defined to satisfy 
\begin{equation}
\big[Q_a(S),T_{(k)}^{q}(S)\big] = \sum_m (Q_{(k),a})_{mq} T_{(k)}^{m}(S) .
\end{equation}
These generalize the commutation relations for the SU(2) tensor operators given in Eq.~\eqref{eq:STalgebra}. We will call the tensor operator representation ``irreducible'' if $(k)$ corresponds to an irreducible representation, and we will work exclusively with irreducible representations and tensor operator representations of compact groups.\par
For states, irreducible representations on a space $S_1 \otimes S_2$  can be built out of tensor product representations of states in spaces in $S_1$ and $S_2$ with Clebsch-Gordan coefficients $C$:
\begin{equation}
\begin{aligned}
    |(k), q\rangle(S_1 \otimes S_2) = \sum_{m_1 m_2 }& C^{(k),q}_{(k_1), m_1, (k_2), m_2}(S_1, S_2)  \\
    &\times|(k_1), m_1\rangle(S_1) |(k_2), m_2\rangle(S_2)
\end{aligned}
\end{equation}
The similar transformation properties of states and tensor operators mean that for the same $k$'s and $q$'s, tensor operators can be built using the \textit{same} Clebsch-Gordan coefficients:
\begin{equation}
\begin{aligned}
    T_{(k)}^{q}(S_1 \otimes S_2) = \sum_{m_1 m_2 } & C^{(k),q}_{(k_1), m_1, (k_2), m_2} (S_1, S_2) \\
    & T_{(k_1)}^{m_1}(S_1) T_{(k_2)}^{m_2}(S_2)
\end{aligned}
\end{equation}
which reduces to Eq.~\eqref{eq:STalgebra} for spherical tensors. Furthermore, the types of irreducible tensor representations appearing within the tensor product of tensor operators are the same as the irreducible representations appearing in the Clebsch-Gordan decomposition of tensor products of group representations.

From the above, if one can find a complete set of tensor operators on a single site ${i}$, one can construct a complete basis of multi-site tensor operators through the use of Clebsch-Gordan coefficients. One procedure to find a tensor operator basis (single-site or otherwise, but we will apply it to a single site) is, in analogy to states, to start with a ``highest weight'' operator and ``lower'' it through iterated commutations with the lowering operators $Q^-_i(S)$.

This procedure for making a single-site basis through iterated commutators highlights another strong connection between representations and tensor operator representations. Call the representation of the group on a single site $R$ (i.e. $R$ is generated by $Q_a(1)$ with $S=1$ referring to a single site. We will omit $(S)$). This representation need not be irreducible nor fundamental. For ease, we will assume we are working in a basis where the raising and lowering operators of this representation are purely real; this can be done for compact groups through a unitary basis transformation. 

Then, consider one of the commutators of the lowering operator(s) with some intermediate tensor operator ${A}$ generated in this procedure: 

\begin{equation}
\begin{aligned}
    &\big[ Q^-(1), \sum_{lr} A_{lr} |l\rangle \langle r| \big]\\
    & =\sum_{lr}A_{lr}\bigg[ Q^-(1)|l\rangle \langle r| - |l\rangle \langle r| Q^-(1)\bigg]\\
    &= \sum_{lr}A_{lr}\bigg[ Q^-(1)|l\rangle \langle r| + |l\rangle \langle r| (-Q^+(1))^\dagger\bigg]
\end{aligned}
\end{equation}
The $A_{lr}$ can all be taken real when working in the basis described above, so it is straightforward to identify the operator above with a state dual to it by exchanging the bra with its dual ket. 
\begin{equation}
\begin{aligned}
    &\big[ Q^-(1), \sum_{lr} A_{lr} |l\rangle \langle r| \big] \\ & \sim \big(Q^-(1)_l \otimes I_r + I_l \otimes (-Q^+(1)_r)\big) \sum_{lr} A_{lr} |l\rangle | r\rangle
\end{aligned}
\end{equation}
The above is written suggestively. Replacing the representation ${R}$'s ${\{Q^{\pm}_{\alpha}\}\rightarrow\{-Q^{\mp}_{\alpha}\}}$ yields the ``conjugate'' representation $\overline{R}$. Thus, the action of $Q^{-}$ on a single-site tensor operator is equivalent to the action of the representation $R \otimes \overline{R}$ on a two-site state. From this, it is easy to show that the formalism can be used to reproduce the spherical tensors in Eq.~\eqref{eq:3jtensor}.

This gives a straightforward way to enumerate a tensor operator basis on a single site. First, find the Clebsch-Gordan coefficients and decomposition of the representation $R \otimes \overline{R}$. Second, replace the ket on the second site with its dual bra, and the result is a tensor operator that transforms like the state does. 

As an example, consider the ladder operator that generates a tower of states for the AKLT spin-1 chain, $Q^+ = \sum_{i=0}^{L-1} (-1)^i (S^+)^2/2$. This ladder operator, along with its Hermitian conjugate $Q^-$, are the ladder operators associated with a reducible representation of SU(2) generated by $Q^x = (Q^++Q^-)/2$, $Q^y = (Q^+-Q^-)/(2i)$, and $Q^z = [Q^+, Q^-]/2$. This representation is reducible on a single-site, for which 
\begin{equation}
\begin{aligned}
Q^x(1) &= \begin{bmatrix}
0 & 0 & 1/2\\
0 & 0 & 0 \\
1/2 & 0 & 0
\end{bmatrix},\quad Q^y(1) = \begin{bmatrix}
0 & 0 & -i/2\\
0 & 0 & 0 \\
i/2 & 0 & 0
\end{bmatrix},\\
Q^z(1) &= \begin{bmatrix}
1/2 & 0 & 0\\
0 & 0 & 0 \\
0 & 0 & -1/2
\end{bmatrix}
\end{aligned}
\end{equation}
i.e. the $\ket{+}$ and $|-\rangle$ degrees of freedom correspond to a spin-1/2, while the $|0\rangle$, which is annihilated by all the generators, corresponds to a spin singlet. Thus, on a single site, the representation $R$ is $1 \oplus 2$ (labeling the representations by their dimension).

To find a complete basis on a single-site using the above procedure we will find the irreps in $(1 \oplus 2)\otimes (\overline{1} \oplus \overline{2})$.
Note that for SU(2), any conjugate representation $\overline{R}$ is unitarily equivalent to the original $R$ (and hence the overbar is usually left off), so the Clebsch-Gordan coefficients will be the usual ones. However, there will be different single-site states corresponding to the highest weight states in the conjugate representation. In particular, as $Q^\pm \rightarrow - Q^\mp$, the highest weight single-site state for $\overline{2}$ will be $|-\rangle$, which will be lowered to $-|+\rangle$. The singlet is identical to itself, so $|0\rangle$ is still the singlet state for $\overline{1}$. 

The resulting irreps in $(1 \oplus 2)\otimes (\overline{1} \oplus \overline{2})$ will be two singlets, two doublets, and one triplet. The tensor operators' explicit forms, found using the regular and conjugate bases above and the usual Clebsch-Gordan coefficients for SU(2), are given in Table~\ref{fig:qtable}. Note in particular that the single site operators $Q^-$, $Q^z$, and $-Q^+$ form a multiplet of tensor operators. This is quite general; for generic nontrivial representations $R$, the generators on a single-site will fill one of the multiplets of tensor operators on that site up to additional phases. This follows from the fact that $R \otimes \overline{R}$ contains the adjoint representation, whose states can be identified with the generators themselves.
\begin{table*}[tb]
  \renewcommand*{\arraystretch}{1.6}
    \centering
    \begin{tabular}{ P{4.5cm}|P{2.5cm}P{2.5cm}P{2.5cm}P{2.5cm}P{2.5cm} }
 \hline
 \hline
& ${q=-1}$ &  ${q=-1/2}$& ${q=0}$ & ${q=1/2}$& ${q=1}$\\
 \hline
 ${2\otimes\overline{2}}$ triplet & ${\ket{-}\bra{+}}$&&${\frac{\ket{+}\bra{+}-\ket{-}\bra{-}}{\sqrt{2}}}$&&${-\ket{+}\bra{-}}$\\
  ${2\otimes\overline{2}}$ singlet &  & & ${\frac{\ket{+}\bra{+}-\ket{-}\bra{-}}{\sqrt{2}}}$& &\\
  ${1\otimes\overline{2}}$ doublet &  & ${\ket{0}\bra{+}}$& &${-\ket{0}\bra{-}}$ &\\
  ${2\otimes\overline{1}}$ doublet &  & ${\ket{-}\bra{0}}$ & & ${\ket{+}\bra{0}}$&\\
 ${1\otimes\overline{1}}$ singlet & & &${\ket{0}\bra{0}}$  & &\\
 \hline 
 \hline
    \end{tabular}
    \caption{A complete basis of tensor operators for single-site $Q$-SU(2) formed from the irreducible representations in the Clebsch-Gordan decomposition of $(2 \oplus 1) \otimes (\overline{2} \oplus \overline{1})$. For SU(2), as noted in the text, any conjugate representation $\overline{R}$ is unitarily equivalent to the original $R$ and hence the bar is usually left off, but we keep it here for clarity. The tensor operators are organized according to dimension and the amount they change the $Q^z$ magnetization.}
    \label{fig:qtable}
\end{table*}

From this complete single-site basis, we can construct complete bases of irreducible tensor operators on multiple sites by using Clebsch-Gordan coefficients. We wrote above
\begin{equation}
\begin{aligned}
    T_{(k)}^{q}(S_1 \otimes S_2) = \sum_{m_1 m_2 } & C^{(k),q}_{(k_1), m_1, (k_2), m_2} (S_1, S_2) \\
    & T_{(k_1)}^{m_1}(S_1) T_{(k_2)}^{m_2}(S_2)
\end{aligned}
\end{equation}
Here, specifying the sites $S_1$ and $S_2$ is useful for handling generators that ``have momentum'', such as $Q^+ = \sum_{i} e^{i p r_i} Q^+_i$. The only difference between $C^{(k),q}_{(k_1), m_1, (k_2), m_2} (S_1, S_2)$ and the usual Clebsch-Gordan coefficients is an additional phase factor stemming from $e^{i p r_i}$. Explicitly, building up tensor operators on $M$ contiguous sites by constructing the tensor product representation of $M-1$ sites and the next site will yield a Clebsch-Gordan sum 
\begin{equation}\label{eq:qpcg}
\begin{aligned}
    T_{(k)}^{q}\bigg(\bigotimes_{i=1}^{M}S_{i}\bigg) = \sum_{m_1 m_2} & C^{(k),q}_{(k_1), m_1, (k_2), m_2}e^{i p (m_2-k_2)(M-1)} \\ 
    &  T_{(k_1)}^{m_1}(r_1,\cdots,r_{M-1}) T_{(k_2)}^{m_2}(r_{M}).
\end{aligned}
\end{equation}

An alternative construction for handling these generators with phase factors ${e^{ipr_i}}$ or more generically ${e^{i\phi_i}}$ is to use a single-site basis that differs from site to site by the phase $e^{i p m r_i}$ and its generalizations for operators that raise the magnetization by $m$. In this case, there would not be an additional phase factor in the sum; the operators $T(S)$ would contain the phase factors.


\subsection{Magnon scarred states from tensor operators of ${Q}$-SU(2) }
The procedure described above allows us to interpret scarred Hamiltonians in the literature as a linear combination of ${Q}$-SU(2) tensor operators, where the ${Q}$-SU(2) symmetry is associated with the magnon operators used to generate scar towers atop fully polarized states.\par 

We first note that Eq.~\eqref{eq:DMR} is a reproduction of Eq.~D13 in Ref.~\cite{mark__2020}, where the DM term breaks spin-1/2 SU(2) symmetry and annihilates the ${p=0}$ magnon tower ${\ket{\psi_{n}}=(S^+)^n\ket{\text{down}}}$. In the same work, it was shown that the model
\begin{equation}
\begin{aligned}
H&=J_{\text{XXZ}}H_{\text{XXZ}}+J_{z}^{\text{DM1}}H^{\text{DM}}_{1}(z)\\
    H_{\text{XXZ}}&=\sum_{i}S^x_{i}S^x_{i+1}+S^y_iS^y_{i+1}-S^z_iS^z_{i+1}
\end{aligned}
\end{equation}
hosts the ${p_0=\pi}$ magnon scar tower ${\ket{\psi_{n}}=(Q^+(\pi))^n\ket{\text{down}}}$ with ${Q^{\pm}(\pi)=\sum_{i}(-1)^{i}S^{\pm}_{i}}$. These two terms in the Hamiltonian can be interpreted as tensor operators of ${Q}$-SU(2), generated by ${\{Q^+(\pi),Q^-(\pi),S^z\}}$. Since
\begin{equation}
\begin{aligned}
    &S^x_{i}S^x_{i+1}+S^y_iS^y_{i+1}-S^z_iS^z_{i+1}\\
    &=\bigg(\frac{\ket{+-}+\ket{-+}}{\sqrt{2}}\bigg)\bigg(\frac{\bra{+-}+\bra{-+}}{\sqrt{2}}\bigg)
\end{aligned}
\end{equation}
is a projector onto the two-site singlet state under ${Q^{\pm}(\pi)=\sum_{i}(-1)^{i}S^{\pm}_{i}}$, ${H_{\text{XXZ}}}$ is scalar under ${Q}$-SU(2). It can also be shown by directly applying 
\begin{equation}\label{eq:qpicg}
\begin{aligned}
        T_{(k)}^{q}(S_1 \otimes S_2) &= \sum_{m_1 m_2} C^{(k),q}_{(k_1), m_1, (k_2), m_2} (-1)^{m_2}\\
        &T_{(k_1)}^{m_1}(r_1) T_{(k_2)}^{m_2}(r_2)
\end{aligned}
\end{equation}
to construct a two-site scalar from the single-site tensors ${T^{1}_{(1)}(r_i)=-S^+_{i}}$, ${T^{0}_{(1)}(r_i)=\sqrt{2}S^z_{i}}$ and ${T^{-1}_{(1)}(r_i)=S^{-}_{i}}$. Using the same method, we see that ${H^{\text{DM}}_{1}(z)}$ is a ${q=0}$ vector. Note that however, ${H^{\text{DM}}_{1}(x)\pm iH^{\text{DM}}_{1}(y)}$ are rank-2 tensors, thus they do not belong to the same multiplet.\par

Turning to spin-${1}$ models, for $Q^+ =  \sum_{i} (-1)^i (S^+_i)^2/2$, a two-site basis can be built out of the single-site basis via Eq.~\eqref{eq:qpicg}. Such a tensor operator basis will contain the irreducible tensor operator representations within $(3+2+2+1+1) \otimes (3+2+2+1+1)$; that is, one $5$, four $4$'s, nine $3$'s, twelve $2$'s, and nine $1$'s.\par

Consider decomposing the spin-$1$ XY model \cite{schecter_weak_2019} into tensor operators. This model has a bimagnon $p=\pi$ scar tower generated by the action of the raising operator, $Q^+ = 1/2 \sum_{i} (-1)^i (S^+_i)^2$, on the base state ${\bigotimes_{i}\ket{-}_{i}}$. The Hamiltonian 
\begin{equation}\label{eq:HXY}
    H_{\text{XY}} = \sum_{i} \left( S^x_i S^x_{i+1} + S^y_i S^y_{i+1} \right) + \sum_{i} \left( h S^z_i + \delta (S^z_i)^2 \right)
\end{equation}
can be broken down into $Q$-SU(2) tensor operators very simply. 
First, note that from the first row of Table \ref{fig:qtable}, $S_i^z$ is proportional to a rank 1, $q=0$ tensor. 
From the second row, $(S_i^z)^2$ is proportional to a rank 0, $q=0$ tensor. 
Finally, notice that
\begin{equation}
\begin{aligned}
    S^x_i S^x_{i+1} + S^y_i S^y_{i+1} &= (\ket{+-}+\ket{-+})\bra{00} \\
    &+ |0+\rangle\langle+0| + |0-\rangle\langle-0| + \text{h.c.}
\end{aligned}
\end{equation}
Here, $(|+-\rangle+|-+\rangle)(\langle 00|)$ is an outer product of two singlet (under $Q$-SU(2)) states, so it is trivally a rank 0, $q=0$ tensor. 
Further, $|0+\rangle\langle+0| + |0-\rangle\langle-0|$ is a rank 1, $q=0$ tensor. It can be found through applying Eq. \eqref{eq:qpicg} to the doublet tensors in the third and fourth rows. 

This decomposition ties into the RSGA-$M$ classification scheme for scarred Hamiltonians introduced in Ref.~\cite{moudgalya_rsga_2020}, c.f Eq.~\eqref{eq:rsgacon}. There, a Hamiltonian $H$ exhibits a restricted spectrum generating algebra of order $M$ (RSGA-$M$) if $M+1$ iterated commutators with $Q^+$ annihilates $H$ while $M$ commutators with $Q^+$ do not.  In the $H_{\text{XY}}$ example above, there are only $k =0, q=0$ and $k =1, q=0$ tensor operators, so by Eq.~\eqref{eq:nestedcom} $H_{\text{XY}}$ exhibits RSGA-$1$.\par 
Likewise, the Hamiltonian (which is the unitary equivalent of ${H_{\text{XY}}}$)
\begin{equation}
    H_{\text{DMI}}= \sum_{i} \left( S^x_i S^y_{i+1} - S^y_i S^x_{i+1} \right) + \sum_{i} \left( h S^z_i + \delta (S^z_i)^2 \right)
\end{equation}
was shown \cite{mark__2020} to host a scar tower with ${p=0}$ bimagnon excitations ${Q^+=\sum_{i}\big(S^+_{i}\big)^2/2}$ on the base state ${\bigotimes_{i}\ket{-}_{i}}$. ${S^z_i}$ and ${(S^z_i)^2}$ are still ${k=1,q=0}$ and ${k=0,q=0}$ tensors of ${Q}$-SU(2). Expressing the DM term as
\begin{equation}
\begin{aligned}
S^x_i S^y_{i+1} - S^y_i S^x_{i+1}&=i(\ket{+-}-\ket{-+})\bra{00}\\
&+i\ket{+0}\bra{0+}+i\ket{0-}\bra{-0}+\text{h.c.},
\end{aligned}
\end{equation}
we see that ${(\ket{+-}-\ket{-+})\bra{00}}$ is again a rank 0, ${q=0}$ tensor, and  $\ket{+0}\bra{0+} + |0-\rangle\langle-0|$ a rank 1, ${q=0}$ tensor. Hence ${H_{\text{DMI}}}$ exhibits RSGA-1.\par

\subsection{Tensor operators of SU(3)}
\begin{table*}[tb]
  \renewcommand*{\arraystretch}{1.5}
    \centering
    \begin{tabular}{ P{4.5cm}|P{2cm}|P{3.5cm}P{3.5cm}P{3.5cm} }
 \hline
 \hline
       &    & ${q_2=-\sqrt{3/2}}$ & ${q_2=0}$& ${q_2=\sqrt{3/2}}$\\
 \hline
  & ${q_1=-\sqrt{2}}$& & ${\ket{0}\bra{+}}$ & \\
  
 &  ${q_1=-1/\sqrt{2}}$ & ${\ket{-}\bra{+}}$& &${\ket{0}\bra{-}}$\\
 
              \multirow{2}{*}{ ${3\otimes\overline{3}}$ octuplet }& \multirow{2}{*}{ ${q_1=0}$}&  & \multicolumn{1}{c}{${\frac{\ket{+}\bra{+}-\ket{0}\bra{0}}{\sqrt{2}}}$} \\
                             &   & & \multicolumn{1}{c}{${-\frac{\ket{+}\bra{+}+\ket{0}\bra{0}-2\ket{-}\bra{-}}{\sqrt{6}}}$} \\
   &   ${q_1=1/\sqrt{2}}$ & ${-\ket{-}\bra{0}}$& &${\ket{+}\bra{-}}$\\
   
      &   ${q_1=\sqrt{2}}$ & & ${-\ket{+}\bra{0}}$&\\
   
${3\otimes\overline{3}}$ singlet  & ${q_1=0}$& &${\frac{\ket{+}\bra{+}+\ket{0}\bra{0}+\ket{-}\bra{-}}{\sqrt{3}}}$  &\\
 \hline 
 \hline
    \end{tabular}
    \caption{A complete basis of tensor operators for single-site SU(3) formed from the irreducible representations in the Clebsch-Gordan decomposition of $3 \otimes \overline{3}$. The tensor operators are organized according to dimension and the amount they change the eigenvalues of the Cartan subalgebra.}
    \label{fig:SU3table}
\end{table*}
As another brief example of the procedure described in Appendix~\ref{sec:constr}, we show how to construct the tensor operator representations of SU(3) on a single $3$ site.

We will label the $3$ representation on a single site through $|+\rangle$, $|0\rangle$, $|-\rangle$, ordered from the highest to the lowest weight states. There are three raising operators, 
\begin{equation}
\begin{aligned}
Q^+_1 &= \begin{bmatrix} 
0 & 0 & 0\\
0 & 0 & 1 \\
0 &  0 & 0
\end{bmatrix},\quad Q^+_2= \begin{bmatrix}
0 & 1 & 0\\
0 & 0 & 0 \\
0 &  0 & 0
\end{bmatrix},\\ 
Q^+_3&= \begin{bmatrix}
0 & 0 & 1\\
0 & 0 & 0 \\
0 &  0 & 0
\end{bmatrix}
\end{aligned}
\end{equation}
which act nontrivially only as follows: $Q^+_1|-\rangle = |0\rangle$, $Q^+_2|0\rangle = |+\rangle$, and $Q^+_3 |-\rangle = |+\rangle$. The corresponding states in the conjugate representation, again ordered from highest to lowest weight, will be $|-\rangle$, $-|0\rangle$, and $|+\rangle$.

SU(3) has a Cartan subalgebra of rank two, which means that states in the representation can be labeled by the eigenvalues of two diagonal operators, 
\begin{equation}
Q_1^z = \frac{1}{\sqrt{2}} \begin{bmatrix}
1 & 0 & 0\\
0 & -1 & 0 \\
0 &  0 & 0
\end{bmatrix},\quad Q_2^z = \frac{1}{\sqrt{6}} \begin{bmatrix}
1 & 0 & 0\\
0 & 1 & 0 \\
0 &  0 & -2
\end{bmatrix}.
\end{equation}
The irreducible representations in the tensor product of $3$ and $\overline{3}$ are $3 \otimes \overline{3} = 1+8$. Using the corresponding Clebsch-Gordan coefficients to construct the states in the representations on the right hand side, and changing the kets to bras, we find Table~\ref{fig:SU3table}.\par  
By inspection, the operators in the table can all be identified as proportional to raising operators, lowering operators, the Cartan subalgebra, or the identity. Thus, this example also illustrates the fact noted above that $R \otimes \overline{R}$ generically contains the adjoint representation and a singlet, which in this case can be identified with the eight generators of SU(3) and the identity operator respectively.\par 

A comment is in order. There can be some ambiguities in the Clebsch-Gordan coefficients for higher-rank groups. SU(2) has a natural basis for each of its irreducible representations, where each state is characterized by its value under $S^z$. This removes any ambiguity from the Clebsch-Gordan coefficients of SU(2). On the other hand, for SU(3), there are irreducible representations where the values in the Cartan subalgebra do not uniquely label the states (i.e. there are weight multiplicities), so there can multiple natural choices of basis and hence different Clebsch-Gordan coefficients. We are using the Clebsch-Gordan coefficients constructed from the method in \cite{alex_numerical_2011}, which makes the tensor operators proportional to the Gell-Mann basis for the generators of SU(3).

%
\bibliographystyle{apsrev4-1}
\bibliography{draft}

\end{document}